\documentclass{article}

\usepackage{arxiv}
\usepackage[utf8]{inputenc} % allow utf-8 input
\usepackage[T1]{fontenc}    % use 8-bit T1 fonts
\usepackage{url}            % simple URL typesetting
\usepackage{booktabs}       % professional-quality tables
\usepackage{amsfonts}       % blackboard math symbols
\usepackage{nicefrac}       % compact symbols for 1/2, etc.
\usepackage{microtype}      % microtypography

\usepackage{amsmath,amssymb}
\usepackage[colorlinks,linkcolor=red,citecolor=blue,urlcolor=blue]{hyperref}

\usepackage{anyfontsize}
\usepackage[table]{xcolor}

\usepackage{array}

\usepackage{bm}

\usepackage{siunitx}

\usepackage{listings}
\newcolumntype{+}{!{\vrule width 2pt}}

\newlength\savedwidth

\newcommand\thickhline{\noalign{\global\savedwidth\arrayrulewidth\global\arrayrulewidth 2pt}%
\hline
\noalign{\global\arrayrulewidth\savedwidth}}

\usepackage[aboveskip=1pt,labelfont=bf,labelsep=period]{caption}

\makeatletter
\renewcommand{\@biblabel}[1]{\quad#1.}
\makeatother

\usepackage{lastpage,fancyhdr,graphicx}
\usepackage{epstopdf}
\fancyheadoffset[L]{2.25in}
\fancyfootoffset[L]{2.25in}
\makeatletter
\newcommand*{\textoverline}[1]{$\overline{\hbox{#1}}\m@th$}
\makeatother

\title{Cross-platform programming model for many-core lattice Boltzmann simulation}

\author{
  Jonas Latt, Christophe Coreixas, and Jo\"{e}l Beny\\
  Computer Science Department, University of Geneva, Carouge, Switzerland\\
  \texttt{jonas.latt@unige.ch}
}

\begin{document}
\maketitle

% Please keep the abstract below 300 words
\begin{abstract}
We present a novel, hardware-agnostic implementation strategy for lattice Boltzmann (LB) simulations, which yields massive performance on homogeneous and heterogeneous many-core platforms. Based solely on C++17 Parallel Algorithms, our approach does not rely on any language extensions, external libraries, vendor-specific code annotations, or pre-compilation steps. Thanks in particular to a recently proposed GPU back-end to C++17 Parallel Algorithms, it is shown that a single code can compile and reach state-of-the-art performance on both many-core CPU and GPU environments for the solution of a given non trivial fluid dynamics problem. The proposed strategy is tested with six different, commonly used implementation schemes to test the performance impact of memory access patterns on different platforms. Nine different LB collision models are included in the tests and exhibit good performance, demonstrating the versatility of our parallel approach. This work shows that it is less than ever necessary to draw a distinction between research and production software, as a concise and generic LB implementation yields performances comparable to those achievable in a hardware specific programming language. The results also highlight the gains of performance achieved by modern many-core CPUs and their apparent capability to narrow the gap with the traditionally massively faster GPU platforms. All code is made available to the community in form of the open-source project \texttt{stlbm}, which serves both as a stand-alone simulation software and as a collection of reusable patterns for the acceleration of pre-existing LB codes.
\end{abstract}

%\linenumbers

% Use "Eq" instead of "Equation" for equation citations.
\section{Introduction}
\subsection{Overview}
A highly challenging aspect of High Performance Computing (HPC) is the need to reformulate and restructure scientific algorithms to perform well on different types of parallel architectures.
In the current hardware landscape, a special focus is devoted to many-core platforms, which include homogeneous systems like the AMD Zen processors investigated in this article, or heterogeneous systems which use a many-core device as an accelerator, including GPUs or Intel's now discontinued Xeon Phi platform. 
Optimized simulation software can be developed for given platforms using device specific programming languages or programming paradigms, including the use of multi-threading on many-core CPUs, or the use of dedicated languages like OpenCL or Cuda on GPUs.
In these cases, the obtained code can be very different from a desirable code for scientific research, as a strong emphasis is put on meeting hardware constraints rather than express the actual numerical model or simulation algorithm. 
Furthermore, this approach requires different code bases to be developed for different architectures, which strongly limits its portability and long-term maintainability.

For this reason, multiple language extensions (such as OpenMP and OpenACC) or completely new programming languages (such as Futhark) have been introduced, which propose to achieve notable performance improvements on parallel hardware in a platform-independent manner, through the application of targeted directives or through the adoption of a specific implementation model. 
In the recent literature, special attention is devoted to functional paradigms, the principles of which are summarized by the so-called map-reduce formalism. 
In this case, an algorithm to be executed on a collection of data elements is expressed in terms of two element access functions, the ``map'' operation that applies a transformation to a data element, and the ``reduce'' operation that expresses a global reduction over the transformed elements. 
The actual, repeated execution of these operations is expressed by the framework instead of the user, allowing for automatic adaptation and optimization of a code on different devices. 

This philosophy of automatic acceleration of functional code is embraced by the C++17 standard. 
It introduces execution policies for standard algorithms, which we will refer to as Parallel Algorithms, that allow parallel and SIMD optimizations. 
Examples of algorithms that are accounted for by execution policies are \texttt{transform-reduce} (the C++ equivalent of map-reduce), \texttt{for\_each}, which only addresses the ``map'' part of map-reduce, or problem-specific functions like the sorting algorithm \texttt{sort}. 
Parallel Algorithms do not necessarily exhibit better performance than traditional language extensions, as a given implementation of the C++ standard library may for example decide to execute Parallel Algorithms with the help of OpenMP directives. 
They do however offer an elegant formalism, as all parallel constructs are expressed as an inherent part of the C++ language. 
They furthermore offer a strong guarantee of cross-platform compatibility and long-term maintainability. Indeed, the C++ language is standardized by the International Organization for Standardization ISO and widely embraced by the scientific community. In addition, the maintainability of a code written in C++ is guaranteed, because the language is known for its success in maintaining backward compatibility over several decades.

While the formalism of Parallel Algorithms is quite recent, it has been tested quite thoroughly on many-core CPU systems thanks to the available implementation through Intel's Threading Building Blocks, or its implementation in Visual C++. 
Only recently though, less than three months before the initial submission of this manuscript, an implementation has been made available that ports Parallel Algorithms to GPUs in form of the NVIDIA HPC SDK C++ compiler, allowing to take profit of a truly cross-platform experience of this formalism.

In this manuscript, we propose a model for the implementation of lattice Boltzmann (LB) codes within the framework of C++ standard algorithms, and show the potential for accelerating such codes with the help of execution policies. 
One of the challenges consists in the implementation of the streaming step, which is non-local and does therefore not fit naturally into the framework of a repeated application of the ``map'' element access function to single elements.
Instead, we take profit from the liberties allowed by the C++ standard to deduce the global index within a data collection for certain types of iterators (including raw pointers), and manipulate these indices further to achieve non-local data accesses. 
While this strategy incurs the cost of added arithmetic operations, the numerical tests show that these do not impact the overall performance in a way that would be prohibitive for the use of this method in a HPC environment.

The proposed modeling environment relies on the usual LB collision-streaming cycles but expresses the local collision through a cell-independent scheme. This approach opposes the frequently adopted strategy of adjusting the local collision scheme to implement boundary conditions (see \textit{e.g.} the Palabos software~\cite{latt_palabos_2020}). 
Instead, the streaming model is locally adjusted to react to encounters with a wall and implement general Dirichlet boundary conditions, following the common, link-based bounce-back approach~\cite{ladd_numerical_1994, kruger_lattice_2017}. 
This approach allows to build a framework with the ability to treat both bulk and boundaries coherently across a large variety of collision models, in a manner that is compatible with efficient many-core parallelism. 
The presented validation case of a lid-driven 3D cavity shows that the selected boundary treatment does not need to hide behind more sophisticated approaches, as the present framework shows to produce stable and accurate results even with the notoriously unstable BGK collision model.
It is pointed out that, although this path is not further explored in this work, the link-based bounce-back collision model can be extended to the treatment of general curved boundary conditions while remaining entirely local (see \textit{e.g.}~\cite{marson_enhanced_2020}).

The result of our efforts is a framework that allows to freely combine six different memory access schemes with nine different LB collision models and bounce-back based Dirichlet walls. In all cases, efficient, platform agnostic many-core parallelism is available. The framework is disseminated in form of the open-source C++ library \texttt{stlbm}, which offers an extensive collection of reusable codes and patterns for the acceleration of LB software. To enhance the educational value of the \texttt{stlbm} library, selected self-contained single-file codes are also provided, which reduce the full software framework to a single memory access scheme and LB collision model. They illustrate that it is possible to write a cross-platform, efficiently parallelized LB application in a readable manner and with less than 500 lines of code, including all necessary data analysis and post-processing functionalities.

The remainder of this introductory section provides a short overview of programming models used to implement parallel software on many-core platforms, and specifically on strategies used to accelerate LB programs on such platforms. %The \nameref{sec:methods} part
Section~\ref{sec:methods} summarizes the method used in this article to implement the LB algorithm and explains in some detail the six schemes (with different data structures and memory access patterns) and nine collision models available in \texttt{stlbm}. It then explains the programming model of C++ Parallel Algorithms, and explains the methodology used to fit the LB algorithm in this model. It finally lists the hardware platforms used to test the performance of the \texttt{stlbm} library. %The Results part 
Section~\ref{sec:results} presents a validation case for the \texttt{stlbm} library (a flow in a lid-driven cavity) and performance measurements for the various implementation schemes, collision models, and hardware platforms considered in this article. Finally, conclusions are drawn in Section~\ref{sec:conclusion}. Section~\ref{sec:supporting_information} provides a link to the repository containing the open-source \texttt{stlbm} library and to the detailed performance measurements obtained with \texttt{stlbm}.

\subsection{Programming models for many-core platforms}
The present article investigates two types of many-core platforms, namely homogeneous systems, in the form of many-core CPUs, and heterogeneous systems, in the form of systems accelerated with GPUs. 
In the latter case, the term ``heterogenous'' refers to the fact that the program is initially executed on a host (the CPU) which handles the program setup and data initialization, after which the computationally intensive tasks are typically offloaded to a large extent to a device (the GPU). Other types of devices, which are not explicitly reviewed here, include FPGAs and Neural Processing Units (NPUs).

Applications on many-core platforms can be programmed using low-level programming models, which include threading mechanisms such as POSIX Threads~\cite{alfieri_efficient_1994} for homogeneous systems. On heterogeneous systems, a strong standard is available in the form of OpenCL, which offers a vendor-independent programming model. 
In practice however, the vendor specific Cuda programming language has established itself as a \textit{de facto} standard for heterogeneous GPU platforms~\footnote{NVIDIA CUDA Toolkit. \url{https://developer.nvidia.com/cuda-toolkit}}, a fact which is sometimes attributed to the difficulty of achieving optimal cross-vendor performances in view of the diversity of many-core architectures~\cite{fang_parallel_2020}.

Higher-level programming models, which are the focus of this article, relieve the programmers from hardware specific details and allow an increased focus on the implemented algorithm, and potentially a greater independence from the selected hardware platform. The two reviewed approaches are
\begin{description}
\item[- Directive based programming models.] This approach, based on extra-language directive annotations, includes industry standards like OpenMP~\footnote{The OpenMP API specification for parallel programming. 
\url{https://www.openmp.org/}}, which is mostly targeted at homogeneous many-core systems, and OpenACC~\footnote{The OpenACC API specification for parallel programming. 
\url{https://www.openacc.org/}} for heterogeneous platforms. 
They typically allow a code to run both on conventional platforms (on which the directives are simply ignored) and on accelerator-supplied systems.

\item[- Programming models based on C++ and on the C++ STL.] C++-based programming models include DPC++ that is part of Intel's OneAPI~\footnote{Intel’s OneAPI. \url{https://software.intel.com/en-us/oneapi}}, C++ AMP which is based on DirectX11, AMD's GPU-oriented C++ API~\footnote{HCC: An open source {C++} compiler for heterogeneous devices. \url{https://github.com/RadeonOpenCompute/hcc}}, and the unified PACXX programming model~\cite{haidl_pacxx_2014} that allows to write both host and device code within the C++14 standard. 
Recently, an increased interest is observed for programming models that are based on the C++ Standard Template Library (STL) or on extensions thereof. 
For example, Thrust~\cite{bell_thrust_2012} is a parallel template library in C++ inspired by the STL, which allows a hardware independent programming style for both many-core CPUs and Nvidia GPUs. 
However, hardware independent program development for many-core platforms is also possible within the strict limits of the C++17 standard, which introduces the concept of Parallel Algorithms. 
This concept is exploited by Intel's Threading Building Blocks (TBB)~\cite{kim_multicore_2011}, as well as by the STL implementation of Visual C++, which provide a backend to C++ Parallel Algorithms for homogeneous many-core systems. 
Finally, NVidia introduced very recently a version of the NVidia HPC SDK which offers an NVidia GPU backend to C++ Parallel Algorithms. 
The TBB and the NVidia HPC SDK backends are used in the present work to test the performances of the proposed programming model for LB applications.
\end{description}

A more thorough review of programming models for homogeneous and heterogeneous many-core platforms is found in~\cite{fang_parallel_2020}.

\subsection{Acceleration of lattice Boltzmann codes on many-core systems}
The LB method has always been considered an excellent candidate for efficient parallelism, most notably thanks to its convenient separation into a local collision step and a streaming step with non-local memory accesses of limited extent. 
Without investigating the numerous implementations and methodologies in detail, we point out the importance which has been devoted, across various systems and platforms, to the reduction of the memory footprint, the improvement of memory access patterns, and the definition of adequate and efficient data structures. 
Pohl \textit{et al.}~\cite{pohl_optimization_2003} propose an implementation strategy which requires the allocation of a single set of populations only (referred to as a \textit{compressed grid} approach) and executes a collision and streaming step for all cells within a single memory traversal (referred to as a \textit{fused collision-streaming step}).
This approach, which is at the heart of the open-source WaLBerla library~\cite{bauer_walberla_2020}, stands at the beginning of a long list of subsequent, similar LB implementation models. 
We mention in particular the swap algorithm~\cite{mattila_efficient_2007}, which is used in the open-source Palabos library~\cite{latt_palabos_2020} and the AA-pattern~\cite{bailey_accelerating_2009}, which was described specifically in the context of GPU implementations of LB models, but has proved highly successful on CPU systems as well~\cite{mohrhard_auto-vectorization_2019}.

With the advent of general-purpose GPU (GPGPU) computing, GPUs have quickly become a target for LB implementations as well. 
An original proposition published in~\cite{ryoo_optimization_2008} proved promising in spite of its fairly poor performance, and was further improved in form of the 2D LB code proposed in~\cite{tolke_implementation_2010}. 
Further performance improvements were exhibited over the years, often by fine-tuning memory access patterns, as published for example in~\cite{kuznik_lbm_2010, obrecht_new_2011, mawson_memory_2014, tran_performance_2017}.

To a lesser extent, LB implementations using the cross-platform formalism OpenCL are found in the literature, which document attempts to target GPUs of different vendors, but also the now discontinued Xeon Phi architecture of Intel (see for example~\cite{mcIntosh-smith_evaluation_2014, obrecht_performance_2015}). 
We also point out the attempts performed with the open-source Sailfish project~\cite{januszewski_sailfish_2014} to use meta-programming capabilities of Python to target multiple platforms through both a Cuda and an OpenCL backend.

\section{Materials and methods}\label{sec:methods}
\subsection{LB algorithm}
The Lattice Boltzmann method is a very specific (physical and numerical) discretization of the Boltzmann equation that splits explicit time iteration into a collision and a streaming step.
Space is subdivided into $N_{tot}$ cells that are in principle equally spaced with a distance $\delta x$ along the three principal Euclidean directions. 
Inhomogeneous cell arrangements can be achieved through mesh refinement strategies~\cite{lagrava_advances_2012, latt_palabos_2020, ASTOUL_JCP_418_2020}, which are however not taken into account in this work. 
The state of the system is defined on every cell by a certain amount of scalar values $f_k (k=0,\ldots,Q-1$), hereafter called populations. 
Their number $Q$ depends on the stencil chosen to discretize the phase space (i.e., physical discretization of the Boltzmann equation).
The present work is based on the commonly used 19-velocity stencil D3Q19~\cite{succi_lattice_2001, chopard_cellular_2012, kruger_lattice_2017}, a choice that limits the current scope of the implementation to isothermal flows. 
A time iteration takes the populations from their state at time $t$ to the next state $t + \delta t$, where $\delta t$ is a constant discrete time step:
\begin{align}
    &\textit{Collision:} & f_k^\mathrm{out}({\bm x}, t) = \Omega_k\left(f({\bm x}, t)\right) \label{eq:collision_step}\\
    &\textit{Streaming:} & f_k({\bm x} + {\bm c}_k \delta t, t + \delta t) = f_k^\mathrm{out}({\bm x}, t)
\end{align}
The 19 discrete velocities ${\bm c}_k$, which connect lattice nodes with near neighbors in lattice units (the components of the ${\bm c}_k$ have integer values), are defined for example in~\cite{kruger_lattice_2017}. 
The temporary values $f_i^\mathrm{out}$, sometimes called ``outgoing populations'', may be stored either in a separate cell array, or in the same array as the $f_k$, using an in-place value replacement scheme (see Section~\ref{sec:data-structure}).

With the link-wise bounce-back scheme used presently to implement Dirichlet boundaries, populations that encounter a physical wall on the way from the original cell to the neighboring one, revert direction and are assigned to the original cell, in opposite direction~\cite{LADD_JFM_271_1994a,kruger_lattice_2017}. The adapted streaming step reads:
\begin{equation}\label{eq:momentum-exchange}
f_{\bar{k}}({\bm x}, t + \delta t) = f_k^\mathrm{out}({\bm x}, t) \underbrace{- 6\,t_k\,\rho_w\,{\bm c}_k\cdot{\bm u}_w}_{\text{momentum exchange}}.
\end{equation}
Here, $\bar{k}$ stands for the opposite direction of $k$ (i.e., ${\bm c}_{\bar{k}}=-{\bm c}_{k}$). In addition, $\bm{u}_w$ is the velocity imposed %enforced
by the Dirichlet boundary condition, and $\rho_w$ is the ``wall density''. 
In the present work, $\rho_w=1$ because the flow runs in a quasi-incompressible regime and exhibits density fluctuations around $1$. 
This boundary condition scheme places the wall half-way between a fluid node and the next solid node, although the exact wall location may be viscosity-depend for low Reynolds number flows. 
This problem is circumvented through the use of the TRT collision term~\cite{ginzburg_two-relaxation-time_2008} (one of the nine models implemented in this article), which should be the preferred choice in pore-scale porous media flows and other high-viscosity flows that are sensitive to the wall location.

The link-wise bounce-back~(\ref{eq:momentum-exchange}) scheme only involves populations that originate from, or move toward, fluid cells. 
Therefore, no storage of populations is required on solid cells. 
The strategy chosen in \texttt{stlbm} to implement boundary conditions consists in allocating nevertheless a layer of wall cells in the solid domain, with 19 storage locations like common fluid cells. 
These wall cells do not store populations, but are rather assigned pre-computed values of the momentum exchange term appearing in Eq~(\ref{eq:momentum-exchange}). 
Thus, the 19 storage locations \texttt{pop\textsubscript{k}} of a cell are utilized as follows:
\begin{align}
\texttt{pop\textsubscript{k}} &= f_k \qquad\text{on fluid cells in pre-collision state (except for the AA-pattern),}\label{eq:storage1}\\
\texttt{pop\textsubscript{k}} &= - 6\,t_k\,\rho_w\,{\bm c}_k\cdot{\bm u}_w \qquad\text{on solid cells.}\label{eq:storage2}
\end{align}
As explained below, the situation is slightly more complex with one of the three implemented memory access schemes, the AA-pattern, which obeys Eq~(\ref{eq:storage1}) at even time steps only, but retrieves pre-collision populations from neighboring fluid cells at odd steps instead. 
With the compact storage scheme expressed in Eqs~(\ref{eq:storage1}) and~(\ref{eq:storage2}), the streaming step is expressed in a uniform manner as an access of neighboring cells to achieve two goals, namely, (1) to copy populations and execute conventional streaming, and (2) to read the pre-computed momentum exchange term and sum it up with a reverted population. 
This latter possibility is expressed in the rightmost column of Table~\ref{tab:memory-access-patterns}.

Implementations of the LB algorithm can be found in the literature using both single-precision and double-precision floating point values. 
Single-precision implementations are particularly popular on GPUs, which are usually substantially more efficient in single precision, except on few high-end GPUs targeted at HPC platforms (which are not tested in the present work). 
Single-precision computations in LB simulations carry however the risk of round-off errors that are difficult to anticipate, especially in research codes, although mitigation strategies exist (see \textit{e.g.}~\cite{skordos_initial_1993}).
To emphasize its goal, which is to be reused in the context of general-purpose research and production codes, the \texttt{stlbm} library has been developed and tested with double precision variables only (although an adaptation to single precision would be easy).

\subsection{LB collision models \label{subsec:coll_models}}
In the present work, nine collision models are considered to evaluate the impact of the collision step~(\ref{eq:collision_step}) on the overall performance of the LB code. Every model is written based on the unified framework proposed by Coreixas et al.~\cite{COREIXAS_PRE_100_2019,COREIXAS_RSTA_378_2020}, which is particularly useful to derive efficient formulations that do not rely on a matrix form. 

More precisely, these collision models include standard operators such as the BGK operator with a weighted second-order equilibrium (BGK-W2~\cite{QIAN_EPL_17_1992}), a non-weighted fourth-order equilibrium (BGK-NW4~\cite{COREIXAS_PRE_100_2019,DEROSIS_ARXIV_2020_01628}), and the two-relaxation-time formulation (TRT~\cite{ginzburg_two-relaxation-time_2008,DHUMIERE_CMA_58_2009}). In addition, multi-relaxation-time collision models based on raw (RM~\cite{DHUMIERE_PAA_159_1992,LALLEMAND_PRE_61_2000,DHUMIERES_TRS_360_2002}), Hermite (HM~\cite{LATT_MCS_72_2006a,SHAN_IJMPC_18_2007,ADHIKARI_PRE_78_2008,CHEN_IJMPC_25_2014}), central (CM~\cite{GEIER_PRE_73_2006,ISHIDA_AIAA_2306_2019,CHAVEZ_Energies_13_2020,DEROSIS_ARXIV_2020_01628}), central Hermite (CHM~\cite{MATTILA_PF_29_2017,SHAN_PRE_100_2019,HOSSEINI_RSTA_378_2020}) moment spaces, as well as cumulants (K~\cite{GEIER_CMA_70_2015,GEHRKE_CF_156_2017,SITOMPUL_JCP_390_2019,NISHIMURA_AIAA_3526_2019}), are further considered. Finally, a multi-relaxation-time formulation of the recursive regularized (RR) collision model is included due to its interesting stability property~\cite{MALASPINAS_ARXIV_2015,COREIXAS_PRE_96_2017,BROGI_JASA_142_2017,JACOB_JT_19_2018,WISSOCQ_PRE_102_2020}. It is worth noting that most of these multi-relaxation time collision models have been proposed for the D2Q9 and D3Q27 lattices, but in this work, we used \textit{non-weighted} D3Q19 formulations that are summarized in Appendix G of Ref.~\cite{COREIXAS_PRE_100_2019}. It was also made possible for the user to freely adjust the bulk viscosity, through an extra parameter, in order to further increase the stability of the code when needed.

More details regarding the implementation of these collision models is provided in Section~\ref{subsec:minimization}, whereas their impact on the performance is presented in Section~\ref{subsubsec:coll_model_perfo}. 

\subsection{Data structure and ordering of populations}\label{sec:data-structure}

To implement an LB scheme, the populations $f_k$ must be stored at a provided memory location. 
This memory location can vary in time, and it may or may not be identical for pre- and post-collision variables ($f_k$ and $f_k^\mathrm{out}$ respectively).
A naive attempt to implement an in-place algorithm, which is to rely on the same storage space for populations at time $t$ and $t+\delta t$, would lead to a conflict with the non-local nature of the streaming step. 
Indeed, as post-collision populations are streamed to neighboring cells, they potentially overwrite populations that have not yet completed their collision-streaming cycle, hence leading to erroneous results. 
In search of a solution, additional constraints to consider are:
\begin{description}
\item[- Reduction of memory requirements.] A low memory footprint is desirable because (1) memory may be a critical resource on a given system, and (2) the total amount of memory processed by a system may impact its performance.
\item[- Reduction of memory accesses.] Although the actual picture is complicated by the presence of cache mechanisms, performance is generally improved by limiting the number of read- and write-accesses to the main memory.
\item[- Thread safety.] This feature is required to allow the multi-threaded execution models used in this article. 
Thread safety is achieved without synchronization primitives or locks, by applying schemes that naturally avoid race conditions. 
A global synchronization is applied only once per time iteration (or twice, in case of the swap algorithm). 
Within this context, achieving thread safety amounts to making sure that the collision-streaming algorithm is independent of the order of traversal of the cells.
\end{description}
This article relies on three popular thread-safe schemes (double-population scheme, swap algorithm, and AA-pattern), which are shortly described below. Their corresponding algorithms are summarized in Table~\ref{tab:memory-access-patterns}.
All three schemes rely on an array of $N_{tot} \times 19$ floating-point variables \texttt{pop}. This amounts to 19 variables for each node location ${\bm x}$, which are accessed through the syntax \texttt{pop\textsubscript{k}(}$\bm x$\texttt{)}. 
The actual memory layout of the \texttt{pop} array is discussed at the end of this section.

\paragraph{Double-population scheme.}
Probably the simplest of the three algorithms, the double-population scheme allocates a second array of size $N_{tot} \times 19$, hereafter called \texttt{out}. 
While this strategy doubles the total memory requirements, it simplifies the algorithm, as both the $f_k(t)$ and $f_k^\mathrm{out}(t)$ values are stored at the memory location \texttt{pop\textsubscript{k}}, while the streamed variables $f_k(t+1)$ are kept separately in \texttt{out\textsubscript{k}}. 
Therefore, the collision and streaming steps can be fused (\textit{i.e.} they are executed within a single memory traversal), as any access conflicts are naturally avoided. After a time iteration, an exchange of the arrays \texttt{pop} and \texttt{out} guarantees that the streamed, temporary populations are reused for the subsequent cycle. 
While the large memory footprint of this scheme may result in severe performance penalties on CPUs (due to a less efficient use of cache memory),
benchmark tests show that it performs quite well on GPUs (see Section~\ref{sec:validation}).
It is also noted that the overall number of memory accesses is kept low in this scheme, with a total of $N_{tot} \times 19$ read accesses and $N_{tot} \times 19$ write accesses per time iteration.

\paragraph{Swap algorithm.} This approach relies on the observation that for every population sent during streaming (from a cell to its neighbor), the cell also receives a population from the same neighbor. 
Therefore, the two copy operations can be fused into a single value swap, which prevents any value from being overwritten. 
Consequently, the algorithm is in-place and relies on no further global memory than the \texttt{pop} array. 
Given that a swap operation includes two copies of the streaming step, the loop executing the streaming operation on a cell spans over nine populations only instead of the usual 18 -- with the understanding that the nine opposite populations are taken care of by the swap operation of a neighboring cell. 
To guarantee that the swapped variables are found in matching locations, post-collision populations must be stored at a storage location associated to the opposite pre-collision population: \texttt{pop\textsubscript{\textoverline{k}}} $\leftarrow$ $f_k^\mathrm{out}$. 
While it is memory efficient, this algorithm can unfortunately not at the same time offer a fused collision-streaming cycle and offer thread safety. 
Indeed, while carrying out a swap operation, it must be guaranteed that the populations of the neighbor involved in the swap are already in post-collision state. 
In a fused collision-streaming scheme, this can be enforced by adjusting the order of traversal of cells with the ordering of the discrete velocities, as it is for example done in the Palabos code~\cite{latt_palabos_2020}. 
The dependence on the order of traversal however violates the principle of threat safety. 
Instead, the version of the swap algorithm used presently splits collision and streaming in two steps requiring each a traversal of the full set of data in \texttt{pop}, and separated by a thread synchronization. 
Thus, the memory needs are half those of the double-population scheme, but the number of memory accesses is doubled, with a total of $2\,N_{tot} \times 19$ read accesses and $2\,N_{tot} \times 19$ write accesses per time iteration.

\paragraph{AA-pattern.} This pattern combines the advantages of the double-population scheme and the swap algorithm, as it offers (1) a fused collision-streaming step, and (2) a single-memory implementation (since it requires no other global data than the \texttt{pop} array).
To achieve this, the data is however stored in different locations at two subsequent time step, and the algorithm must distinguish even and odd time steps.
More precisely, the AA-pattern transfers the streaming step performed at the end of an even iteration to the beginning of the next odd iteration.
Hence, two communication steps are incorporated at odd iterations, in terms of a ``Pull'' operation that gathers the populations from the neighbor cells to a local and temporary array for collision, before a ``Push'' operation that eventually writes the post-collision variables back to the same locations at the neighbors. 
Thread safety is guaranteed by virtue of the fact that a given cell (handled by a single thread) accesses the same non-local data during its read (Pull) and write (Push) access. 
Going into more details, at even steps the populations are stored locally, and the value of $f_k({\bm x})$ is available at the location \texttt{pop}\textsubscript{k}$({\bm x})$. Regarding post-collision values, they are also stored locally but at opposite locations, like in the swap algorithm: \texttt{pop\textsubscript{\textoverline{k}}} $\leftarrow$ $f_k^\mathrm{out}$.
At odd steps, the populations are available on neighboring nodes, with $f_k({\bm x})$ stored at \texttt{pop}\textsubscript{\textoverline{k}}$({\bm x} + {\bm c}_k)$. 

All in all, the AA-pattern is a compressed scheme which, like the swap algorithm, cuts the memory requirements of the double-population scheme in half, but favorably maintains the same number of memory accesses, with an average of $N_{tot} \times 19$ read accesses and $N_{tot} \times 19$ write accesses per time iteration. It should however also be mentioned that the AA-pattern is more complex than the two other schemes, and introduces technical difficulties for the maintenance of an LB software framework, because of the separation into even and odd time steps.

\begin{table}[!htbp]
\centering
\caption{\bf The algorithms executed for the three implemented schemes}
\centering\texttt{nb}\textit{: neighbor, }$\leftarrow$\textit{: assignment, }$\leftrightarrow$\textit{: swap}\\
\begin{tabular}{|l+l|l|l|}
\hline
 & \multicolumn{1}{c|}{\textbf{Collision}} & \multicolumn{2}{l|}{\hspace{10ex}\textbf{Communication}}\\
 & & \textbf{\texttt{nb} is fluid} & \textbf{\texttt{nb} is wall}\\\thickhline
&& \multicolumn{2}{c|}{\textit{Streaming}: \texttt{k} runs over 18 values}\\
2Pop &
\texttt{pop\textsubscript{k}} $\leftarrow$ $\Omega$\textsubscript{k}(\texttt{pop})
&
\texttt{nb\_out\textsubscript{k}} $\leftarrow$ \texttt{pop\textsubscript{k}}
&
\texttt{out\textsubscript{\textoverline{k}}} $\leftarrow$ \texttt{pop\textsubscript{k}} + \texttt{nb\_pop\textsubscript{k}}\\\hline
&& \multicolumn{2}{c|}{\textit{Streaming}: \texttt{k} runs over 9 values}\\
Swap &
\texttt{pop\textsubscript{\textoverline{k}}} $\leftarrow$ $\Omega$\textsubscript{k}(\texttt{pop})
&
\texttt{pop\textsubscript{\textoverline{k}}} $\leftrightarrow$ \texttt{nb\_pop\textsubscript{k}}
&
\texttt{pop\textsubscript{\textoverline{k}}} $\leftarrow$ \texttt{pop\textsubscript{\textoverline{k}}} + \texttt{nb\_pop\textsubscript{k}}\\\hline

AA/Even &
\texttt{pop\textsubscript{\textoverline{k}}} $\leftarrow$ $\Omega$\textsubscript{k}(\texttt{pop})
& \multicolumn{2}{c|}{\textit{Pull}: \texttt{k} runs over 18 values}\\\cline{1-2}

AA/Odd &&
\texttt{tmp\textsubscript{k}} $\leftarrow$ \texttt{nb\_pop\textsubscript{\textoverline{k}}}
&
\texttt{tmp\textsubscript{k}} $\leftarrow$ \texttt{pop\textsubscript{k}} + \texttt{nb\_pop\textsubscript{\textoverline{k}}}\\
& \texttt{tmp\textsubscript{k}} $\leftarrow$ $\Omega$\textsubscript{k}(\texttt{tmp})
& \multicolumn{2}{c|}{\textit{Push}: \texttt{k} runs over 18 values}\\
&& \texttt{pop\textsubscript{k}} $\leftarrow$ \texttt{tmp\textsubscript{k}}
&
\texttt{pop\textsubscript{\textoverline{k}}} $\leftarrow$ \texttt{tmp\textsubscript{k}} + \texttt{nb\_pop\textsubscript{k}}\\\hline
\end{tabular}
\caption*{This table summarizes the algorithm that must be executed on all fluid cells in each of the three schemes. Only in the swap algorithm, wall cells are also included in the collision-streaming cycle, and apply the following operation to all fluid cells among the $9$ considered neighbors: \texttt{nb\_pop\textsubscript{k}} $\leftarrow$ \texttt{pop\textsubscript{\textoverline{k}}}. The following notation is used: \texttt{pop} denotes the 19 variables of the local cell, at position $\bm{x}$, and \texttt{nb\_pop} the variables of the neighbor cell in direction \texttt{k}, at position $\bm{x}$ + $\bm{c}$\textsubscript{k} (\textit{e.g.} \texttt{nb\_pop}\textsubscript{\textoverline{k}} denotes the variable of index \texttt{\textoverline{k}} of the cell at position $\bm{x}$ + $\bm{c}$\textsubscript{k}). The double-population scheme uses a duplicate array of cells which are denoted as \texttt{out}. Finally, the AA-pattern uses a temporary cell \texttt{tmp} which is entirely local and does not require the allocation of a global cell-array like \texttt{out}.}

\label{tab:memory-access-patterns}
\end{table}

\paragraph{}Finally, the data can be aligned in two different ways in memory, independently of the chosen data structure. 
In the first layout, referred to as array-of-structure (aos), the populations of a given cell are aligned consecutively in memory. 
In the second approach, which carries the name of structure-of-array (soa), all populations corresponding to a given direction k are consecutive. 
The two following statements summarize the essence of the two data layouts using a C-array syntax:
\begin{align*}
    &\text{Array-of-structure:} & \text{\lstinline{double pop[Ntot][19];}}\\
    &\text{Structure-of-array:} & \text{\lstinline{double pop[19][Ntot];}}
\end{align*}
By combining these two data layouts with the three proposed LB implementation schemes, the \texttt{stlbm} library offers a total of six implementation strategies for each collision model, which produce identical results but exhibit different, platform-dependent performance figures, as explored in this article.

\subsection{STL and Parallel Algorithms}
The C++ Standard Template Library (STL) supports a functional programming style, in which algorithms are supplemented with element access functions which customize their behavior. 
The algorithms apply these functions repeatedly to the elements of a data container, either through function pointers or function objects. 
At the application programmer level, an explicit loop is then replaced by a single call to an STL algorithm, hence leading to a programming style that is sometimes considered to be more expressive. 
The machine codes generated by an algorithm invocation or an explicit loop do not necessarily differ, as a compiler can take advantage of the mechanisms of C++ templates to translate the algorithm to an identical or similar loop as the hand-written one. 
This is especially true when the implementation of the element access function is known by the compiler at the time of the algorithm instantiation, in which case the function body can be efficiently inlined.

The following C++ code extract illustrates this concept with the STL algorithm \lstinline{for_each}, used to compute macroscopic variables in the manner of a typical C++ program:

\begin{lstlisting}
using namespace std;
struct Cell {
    array<double, 19> f;
    double rho;
    array<double, 3> u;
};
vector<Cell> cells( nCells );
array<int, 3>* c = ...; // Lattice velocities

// ... Initialize data, start principal time loop.

for_each( execution::par_unseq, begin(cells), end(cells),
          [c](Cell const& cell )
{
    cell.rho = 0.;
    cell.u = {0., 0., 0.};
    for (int k = 0; k < 19; ++k) {
        cell.rho += cell.f[k];
        cell.u[0] += cell.f[k] * c[k][0];
        cell.u[1] += cell.f[k] * c[k][1];
        cell.u[2] += cell.f[k] * c[k][2];
    }
});
\end{lstlisting}
In this example, the data layout follows the principles of an \emph{array-of-structure} (aos), as all variables related to a given cell are located at consecutive memory addresses.
The element access function provided to the \lstinline{for_each} algorithm takes in this case the form of a lambda expression (a \emph{lambda} for short), which defines an anonymous function object. 
A lambda can obtain access to variables from the surrounding scope, in a manner specified in the capture clause. 
In this example, the clause \lstinline{[c]} indicates that the variable \lstinline{c}, which refers to the discrete velocities, is captured by value. For reasons linked to limitations of the GPU backend of Parallel Algorithms, as explained below, the \texttt{stlbm} library defines \lstinline{[c]} as a raw pointer to a heap array.

Finally, the code extract makes use of a parallel version of the \lstinline{for_each} algorithm. Parallel algorithms are available starting with the C++17 standard. They receive an \emph{execution policy} as an additional argument to indicate a parallelization strategy they may use to accelerate the execution of the algorithm. 
The four possible strategies are
\begin{description}
    \item[Sequenced.] Requires that the algorithm's execution may not be parallelized and that the calls to the element access functions must be sequenced. The result is usually the same as in a non-parallel version of the algorithm. The policy is usually invoked through \lstinline{execution::seq}.
    \item[Parallel.] Indicates that the algorithm's execution may be parallelized using multiple threads. The policy is usually invoked through \lstinline{execution::par}.
    \item[Unsequenced.] Provided since C++20 (but available within the C++17 standard in all tested compilers), the policy indicates that the algorithm's execution may be vectorized, by using for example instructions that operate on multiple data items. The policy is usually invoked through \lstinline{execution::unseq}.
    \item[Parallel and Unsequenced.] Indicates that the algorithm's execution may be parallelized in a multi-threaded fashion, and that vectorization is allowed within each thread. The policy is usually invoked through \lstinline{execution::par_unseq}.
\end{description}
In this article, only the \lstinline{execution::par_unseq} policy has been applied, which allowed to achieve best performance on both CPUs and GPUs.

The methodology proposed by C++ Parallel Algorithms faces a challenge on heterogeneous systems. While the memories of the host and the device are often physically distinct on such systems, the C++ standard does not provide any means to transfer data between different memories. The Parallel Algorithms backend for NVidia GPUs circumvents this problem thanks to a memory model called \emph{CUDA Unified Memory}, which provides a single memory address space to access data from both the host and the device. Through support in both the CUDA device driver and the NVidia GPU hardware, a Unified Memory manager automatically transfers data between the two physical memories based on usage. The model is extremely convenient at the user programmer level who entirely avoids explicit data transfers, yet needs to remain aware of the cost of implicit transfers. Current versions of CUDA Unified Memory are however limited to sharing data allocated on the heap, and cannot share stack memory. On NVidia GPU systems, pointers provided to the lambda capture of a parallel algorithm, or variables captured by reference, must necessarily reference heap data. For this reason, all shared data, including the lattice constants, is allocated on the heap in the \texttt{stlbm} library.

While the above code listing efficiently parallelizes the computation of macroscopic variables (which is a fully local step), the same strategy cannot be applied directly to non-local aspects of the LB algorithm. 
Indeed, STL algorithms provide only limited support to express an interaction between container elements. 
This is done either through highly specialized algorithms, such as the parallelized sorting algorithm, or in terms of reduction operations, implemented for example in the \lstinline{transform_reduce} algorithm.
None of these are sufficient to tackle the two following problems that are addressed in this manuscript:
\begin{enumerate}
    \item Implementation of a structure-of-array (soa) data layout. In this case, the populations needed for a collision are no longer consecutive in memory, and a cell can therefore no longer be treated as a single element of a C++ container.
    \item Implementation of the streaming step, which requires access to nearby spatial neighbors.
\end{enumerate}
STL algorithms do however not prohibit neighbor access from being implemented manually, as the full data container can be disclosed to an element access function through the capture.
This strategy is further explained in the following section.

\subsection{Parallel LB algorithm}
The code extract of the previous section uses an anonymous lambda to express the element access function of an STL algorithm, following standard practices in simple STL usage scenarios. 
This strategy however forces the access function to be implemented at its point of usage, which strongly limits the possibility to properly organize a larger code base. 
In the following, function objects are therefore %rather 
instantiated from named classes, both to encourage the reuse of the proposed code in other projects, and to avoid misunderstandings linked to the implicit syntax of lambda capture clauses.

The encapsulation of element access functions in named classes opens the possibility for their instantiation (the function object) to be named and obtain an extended lifetime. 
It is even possible for the data such as the LB populations to be owned and managed by the same object, in which case the LB data and algorithms are unified under the same abstraction. 
In the present manuscript, we prefer however to treat these objects as stateless algorithms, and to manage data in a different scope. 
The purpose of this approach is again the ease of reuse of the proposed codes which, in a considered frequent use scenario, should be applied to accelerate selected parts of an existing code with a pre-existing memory management strategy. 
This LB algorithm class adopts the following canonical shape:

\begin{lstlisting}
struct LBM {                                  // (1)
    using CellData = array<double, 19>;       // (2)
    CellData* lattice;                        // (3)
    array<int, 3>* c;  // lattice velocities  // (4)
    double* t;         // lattice weights
    double omega;      // relaxation parameter
    Dim dim;           // nx x ny x nz dimensions
    
    void operator() (CellData& cell) {        // (5)
        // Implement collision-streaming for a cell.
    }
};
\end{lstlisting}
The enumerated lines require the following clarifications:
\begin{description}
\item[(1)] The class is declared with the \lstinline{struct} keyword to provide public access to all data and methods, as data encapsulation or any other elements of object-oriented programming are outside the scope of the presented work.
\item[(2)] \lstinline{CellData} names the type of a single data element provided to the \lstinline{for_each} algorithm, which must be chosen according to the data layout. Here, the example of an aos layout with consequent cell data is chosen.
\item[(3)] The class variables of the five following lines are the articulated counterpart of a lambda capture. A pointer to the full lattice is captured to allow access to the neighboring cells. In the case of the soa data alignment, the full lattice is also required to gather all populations of the current cell.
\item[(4)] We chose to capture the lattice constants (velocities and weights) by reference and the relaxation time and lattice dimensions by value.
\item[(5)] To turn class instances into a function object, the function call operator is overloaded to implement a collision-streaming cycle for a single mesh grid cell.
\end{description}

A canonical code for the application of an \lstinline{LBM} instance to an LB lattice is expressed as follows:
\begin{lstlisting}
vector<LBM::CellData> lattice_vect(num_elements);    // (1)
LBM::CellData* lattice = &lattice_container[0];      // (2)
LBM lbm{lattice, &c[0], &t[0], omega, {nx, ny, nz}}; // (3)
for_each( execution::par_unseq,
          lattice, lattice + num_elements, lbm );
\end{lstlisting}
This extract is again commented at the enumerated lines:
\begin{description}
\item[(1)] The number of elements equals the number of cells (aos) or 19 times the number of cells (soa).
\item[(2)] The vector is simply used here for automatic management of heap data. The data can also be allocated with the \lstinline{new} operator and deleted manually.
\item[(3)] Just as in a lambda capture, any variable visible in the present scope can be provided to initialize a \lstinline{LBM} object.
\end{description}

Independently of the memory layout strategy (aos or soa),
the data container \lstinline{lattice_vect} is required to allocate a number of floating point variables equal to 19 times the number of cells. 
Differences only appear in their usage pattern. In the algorithmically simpler case of aos, these variables are regrouped by cells, as the type \lstinline{CellData} refers to sequences of 19 floating point variables. 
The \lstinline{for_each} algorithm is then simply applied to the full data space, cell by cell. 
In the soa layout on the other hand, the cell data is non-contiguous in memory, and the \lstinline{CellData} type is chosen to refer to a single floating-point variable. 
The \lstinline{for_each} algorithm can consequently not be applied to the full data space --as this would lead to too frequent invocations of a cell-level collision-streaming cycle-- but spans only over the elements of the first population $f_0$. 
It is the responsibility of the methods of \lstinline{LBM} to compute the indices of the other populations inside the array \lstinline{lattice} and access them manually. 
Similarly, indices are computed to access neighboring cells for the streaming step.

The index of the currently processed cell is not explicitly provided by the \lstinline{for_each} algorithm to its element access function. 
While the index could be reconstructed in a non-parallel version of \lstinline{for_each} by giving up the stateless nature of \lstinline{LBM} and incrementing an internal index at each function call, such an approach would be in obvious violation with a multi-threaded or unsequenced execution of the algorithm. 
Instead, we choose to rely explicitly on the fact that the iterators in this invocation of \lstinline{for_each} are raw pointers, and deduce the index from the difference of memory address between the current and first element of the lattice. 
While it is clear that such an approach could not be extended to other types of C++ containers and iterators, it should also be mentioned that this limitation to the scope of raw data arrays is rather expected in the context of massive HPC. 
It is further pointed out that the proposed code preserves a consistent high-level structure. 
Violations of the safety of data types and memory bounds are limited to a small group of data access function which encapsulate the cases of pointer-based arithmetic operations. 
Consequently, the data types and data access functions are defined as follows in a array-of-structure respectively a structure-of-array layout:
\begin{lstlisting}
// Implementation for array-of-structure layout
struct LBM {
    using CellData = array<double, 19>;
    double& f (int i, int k) {
        return lattice[i][k];
    }
    // ... implementation of collision models
};

// Implementation for structure-of-array layout
struct LBM {
    using CellData = double;
    double& f (int i, int k) {
        return lattice[k * dim.nelem + i];
    }
    // ... implementation of collision models
};
\end{lstlisting}
Based on these access functions, the collision models can then be defined in an identical manner for both data layouts. Furthermore, the strategy for evaluating the index value and manipulating 3D indices is layout independent:
\begin{lstlisting}
struct LBM {
    // ... data definitions
    // Function i_to_xyz transforms a linear index
    // into a Cartesian coordinate triplet
    auto i_to_xyz (int i) {
        int iX = i / (dim.ny * dim.nz);
        int i_yz = i % (dim.ny * dim.nz);
        int iY = i_yz / dim.nz;
        int iZ = i_yz % dim.nz;
        return std::make_tuple(iX, iY, iZ);
    };
    void operator() (CellData& cell) {
        size_t i = &cell - lattice;
        // ... implementation of collision-streaming
    }
}
\end{lstlisting}

\noindent The complete source code of the \texttt{stlbm} can be accessed under the terms of an open-source license under \nameref{S1_Source Code}.

\subsection{Minimization of number of instructions\label{subsec:minimization}}
Two major strategies for improving the efficiency of an HPC code consist in (1) optimizing memory accesses (which we do by testing six different data access strategies), and (2) reducing the number of instructions executed by the code. 
In LB applications, the latter is typically achieved by unrolling the loops over populations, and regrouping and eliminating terms in the resulting arithmetic expressions. In the present work, two versions of the second-order BGK (BGK-W2) and TRT collision models were implemented. 
The first, referred to as the \emph{educational} version, uses only few superficial optimizations and relies on the capabilities of the compiler for further in-depth improvements, in order to encourage readability and re-use of the code. 
The second, referred to as the \emph{unrolled} version uses aggressive loop unrolling and regrouping of terms to achieve manual performance gains.
Interestingly, while the unrolled code versions systematically yield substantially better performance than the educational ones on GPUs, the flagship AMD CPU used in the tests obtained better results with the educational code for the BGK model. 

In the educational code versions, loop unrolling was applied to the computation of the macroscopic variable density and velocity, to eliminate the frequent multiplications with zero-valued components of the discrete velocities $c_i$. 
Furthermore, partial sums that were shared by the expressions for density and velocity were regrouped. 
While the loops for the computation of equilibrium and for executing the collision models were not unrolled, the similarity of the equilibrium term for opposite populations was explicitly exploited. 
For example, as the second-order weighted equilibrium (BGK-W2) for a direction $k$ is written as
\begin{equation}
\mathrm{E}_k = \rho\,t_k\left(1 + 3\,\bm{c}_k\cdot {\bm u} + 4.5\,(\bm{c}_k\cdot {\bm u})^2 - \left|{\bm u}\right|^2\right),
\end{equation}
the equilibrium value in the opposite direction $\bar{k}$, defined by the relation $\bm{c}_{\bar{k}} = - \bm{c}_k$, can be computed from $\mathrm{E}_k$ through the equation

\begin{equation}\label{eq:symmetry_Ek}
\mathrm{E}_{\bar{k}} = \mathrm{E}_k - 6\,\rho\,t_k\,\bm{c}_k\cdot\bm{u}.
\end{equation}

\noindent In the unrolled code, all loops in the collision term were additionally unrolled, and redundant terms eliminated.

Regarding more complex collision models, their implementation relies on the computation of post-collision moments of interest, which are converted back to raw moments (RMs) in order to compute post-collision populations $f_k^{out}$. This methodology allows for a drastic reduction of the number of instructions since it naturally (1) discards multiplication by zero-valued terms, and (2) highlights existing symmetries between opposite post-collision populations. By taking advantage of the latter property, one can reduce the number of instructions required to compute $f^{out}_{\bar{k}}$ by first computing $f^{out}_{k}$ --as already done for $\mathrm{E}_{\bar{k}}$ in Eq~(\ref{eq:symmetry_Ek}). For instance,
\begin{align}
        f^{out}_{0} &= \frac{\rho}{2} (-u_x + M^*_{200} + M^*_{120} + M^*_{102} - M^*_{220} - M^*_{202}),\\
        f^{out}_{10}  &= \frac{\rho}{2} (\phantom{-} u_x + M^*_{200} - M^*_{120} - M^*_{102} - M^*_{220} - M^*_{202}),
\end{align}
with $\bm{c}_0=(-1,0,0)=-\bm{c}_{10}$ based on the velocity ordering convention used in the accompanying codes.
Hence,
\begin{equation}
    f^{out}_{0}  = f^{out}_{10} - \rho (u_x - M^*_{120} - M^*_{102}),
\end{equation}
where $M^*_{pqr}$ ($p,q,r\in \mathbb{N}$) are post-collision RMs defined as 
\begin{equation}\label{eq:RMcoll}
 M^*_{pqr}= (1 - \omega_{pqr})M_{pqr} + \omega_{pqr} M^{eq}_{pqr}.  
\end{equation}
$\omega_{pqr}$ are adjustable relaxation frequencies, and $M^{eq}_{pqr}$ only depend on $\rho$ and $\bm{u}$ (their full expression can be found for all types of moment, e.g., in Ref.~\cite{COREIXAS_PRE_100_2019}). In addition, the computation of RMs
\begin{equation}\label{eq:RM}
M_{pqr}=\sum_k c_{kx}^p c_{ky}^q c_{kz}^r f_k,    
\end{equation}
as well as other types of moments, is of paramount importance since it drastically impacts the performance of the code. As a matter of fact, the performance can be divided by a factor two or three if their general formulation is used. Consequently, aggressive optimizations (loop unrolling and regrouping of terms) were also applied to increase the performance of these collision models. For all models other than BGK-W2 and TRT, no ``educational'' version was produced.

As a sidenote, the multi-relaxation-time formulations considered in this study could be further optimized by limiting their generality in part, e.g. by imposing $\omega_{pqr}=1$ in Eq~(\ref{eq:RMcoll}).
The latter value is commonly used in the LB community for high-order ($p+q+r\geq3$) post-collision moments, as well as those related to bulk viscosity, as it leads to an increased stability for high-Reynolds number flow simulations and/or in under-resolved conditions. 
Such a choice would allow us to discard the computation of the corresponding moment~(\ref{eq:RM}) which remains time consuming despite our optimizations. 
%More quantitative investigations are out of the scope of the present paper. This is why results regarding performance gains induced by the equilibration of high-order moments (and those related to the bulk viscosity) will be presented in a future study.
More quantitative investigations regarding performance gains induced by the equilibration of high-order moments (and those related to the bulk viscosity) will be discussed in a future study.
The interested reader may also refer to Refs.~\cite{DEROSIS_PoF_31_2019,DEROSIS_ARXIV_2020_01628} for a brief discussion of this aspect in  the context of multiphysics flow simulations based on CM-LBMs.

Finally, an extra parameter has been added to all multi-relaxation-time formulations, so that the bulk viscosity can be freely adjusted. This is done by modifying the relaxation of second-order moments as follows~\cite{FEI_PRE_97_2018,COREIXAS_PRE_100_2019}
\begin{align}\label{eq:bulk}
M^*_{200}&= M_{200}  - \omega_{+} (M^{eq}_{200}-M^{eq}_{200})  - \omega_{-} (M^{eq}_{020}-M^{eq}_{020})  - \omega_{-} (M^{eq}_{002}-M^{eq}_{002}),\\
M^*_{020}&= M_{020}  - \omega_{-} (M^{eq}_{200}-M^{eq}_{200})  - \omega_{+} (M^{eq}_{020}-M^{eq}_{020})  - \omega_{-} (M^{eq}_{002}-M^{eq}_{002}),\\
M^*_{002}&= M_{002}  - \omega_{-} (M^{eq}_{200}-M^{eq}_{200})  - \omega_{-} (M^{eq}_{020}-M^{eq}_{020})  - \omega_{+} (M^{eq}_{002}-M^{eq}_{002}), 
\end{align}
where $\omega_{-}= (\omega_{\nu_b} - \omega_{\nu})/3$ and $\omega_{+}= \omega_{-} + \omega_{\nu}$. The kinematic viscosity $\nu$, and its bulk counterpart $\nu_b$, are related to these relaxation parameters through $\nu=(1/\omega_{\nu}-0.5)c_s^2$ and $\nu_b=(1/\omega_{\nu_b}-0.5)c_s^2$. The lattice constant $c_s$ is defined as $c_s=1/\sqrt{3}$~\cite{kruger_lattice_2017}.

\subsection{Hardware platforms and compilers}
The computers used to carry out the benchmark simulations are listed below, along with the compiler and compiler options that were used. 
\begin{center}
\begin{tabular}{+ll+}
\thickhline
\multicolumn{2}{+c+}{\bf AMD Ryzen Threadripper (8 cores)}\\\hline
{\bf Model} & AMD Ryzen Threadripper 1900X 8-Core Processor\\
{\bf Cores} & 8 Cores, 16 Threads (SMT on), 1 Socket\\
{\bf Compiler} & \texttt{CLang 10.0.0-4 / -std=c++17 -O3}\\\thickhline
\multicolumn{2}{+c+}{\bf Intel Xeon (8 cores)}\\\hline
{\bf Model} & Intel Xeon E5-2637 v4 @ 3.50GHz\\
{\bf Cores} & 8 Cores, 8 Threads (Hyperthreading off), 2 Sockets\\
{\bf Compiler} & \texttt{GCC/g++ 9.3.0 / -std=c++17 -O3 -march=native}\\\thickhline
\multicolumn{2}{+c+}{\bf Intel Xeon (48 cores)}\\\hline
{\bf Model} & Intel Xeon Gold 6240R @ 2.40GHz\\
{\bf Cores} & 48 Cores, 48 Threads (Hyperthreading off), 2 Sockets\\
{\bf Compiler} & \texttt{GCC/g++ 9.3.0 / -std=c++17 -O3 -march=native}\\\thickhline
\multicolumn{2}{+c+}{\bf AMD EPYC (64 cores)}\\\hline
{\bf Model} & AMD EPYC 7742 64-Core Processor @ 2.24 GHz\\
{\bf Cores} & 64 Cores, 64 Threads (SMT off), 1 Socket\\
{\bf Compiler} & \texttt{CLang 10.0.1 / -std=c++17 -O3}\\\thickhline
\multicolumn{2}{+c+}{\bf NVIDIA GTX 1080 Ti}\\\hline
{\bf Model} & GP102 [GeForce GTX 1080 Ti]\\
{\bf Compiler} & \texttt{nvc++ 20.7-0 / -stdpar -O1 -std=c++17}\\\thickhline
\multicolumn{2}{+c+}{\bf NVIDIA RTX 2080 Ti}\\\hline
{\bf Model} & TU102 [GeForce RTX 2080 Ti]\\
{\bf Compiler} & \texttt{nvc++ 20.7-0 / -stdpar -O1 -std=c++17}\\\thickhline
\multicolumn{2}{+c+}{\bf NVIDIA V100}\\\hline
{\bf Model} & GV100 [Tesla V100-PCIE-32GB]\\
{\bf Compiler} & \texttt{nvc++ 20.7-0 / -stdpar -O1 -std=c++17}\\\thickhline
\end{tabular}
\end{center}

The compiler and compiler options were selected in informal comparative tests designed to identify the most efficient candidates. Except for the GPU versions compiled with \texttt{nvc++}, all binaries were linked with the \texttt{-ltbb} option to link with Intel's Threading Building Blocks.

\section{Results and discussion\label{sec:results}}

\subsection{Benchmark case and code validation}\label{sec:validation}
\begin{figure}[!hbt]
\centering
\hspace*{\fill}
\includegraphics[width=.47\textwidth]{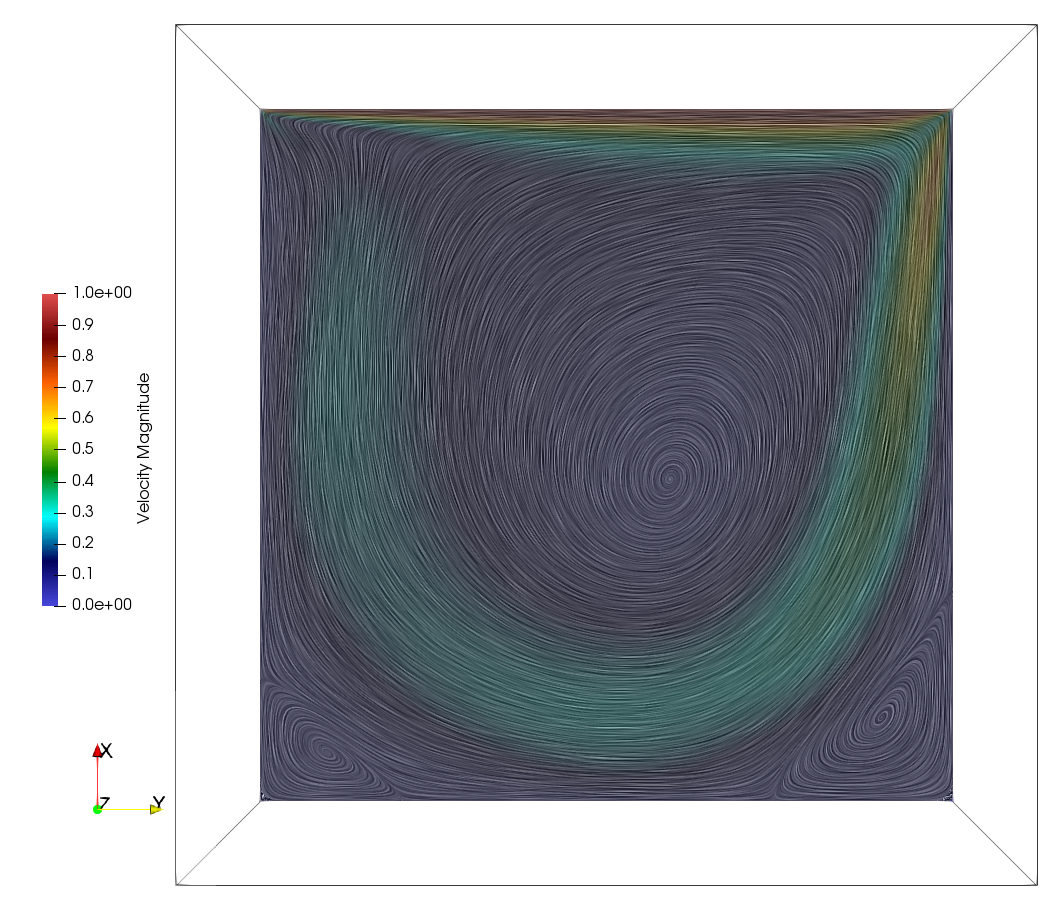}
\hfill
\includegraphics[width=.44\textwidth]{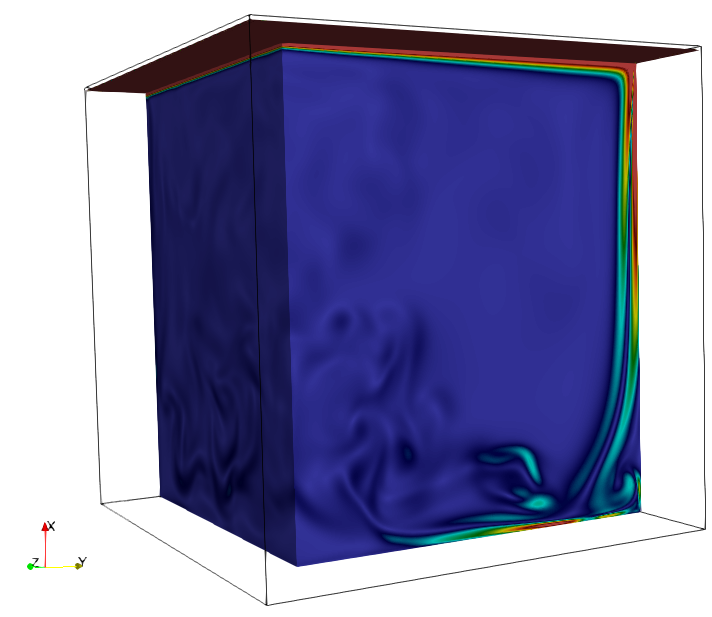}
\hspace*{\fill}\vspace{1ex}

    \caption{{\bf Visualization of simulation results for the lid-driven cavity.} Left image: velocity norm and streamlines of the steady-state results at $\mathrm{Re}=1000$, on the plane defined by $z=0$. Right image: vorticity norm on a selected surface at an instantaneous time step for the fully developed flow at $\mathrm{Re}=10000$.}\label{fig:cavity-velocity-field}
\end{figure}

The code of this article is applied to the benchmark case of a 3D flow in a lid-driven, cubic cavity, under the conditions of a steady, a laminar unsteady, and a turbulent regime. This test is used both to demonstrate the correctness of the code and to explore its efficiency on different hardware platforms depending on the implementation strategy and the LB collision model. This benchmark case, which can be found in numerous articles (see e.g.~\cite{albensoeder_accurate_2005}), is defined as follows. An incompressible, viscous, Newtonian, and homogeneous fluid, the dynamics of which is described by the Navier-Stokes equations, is enclosed in a cubic domain. The 3D fluid-domain is defined by the open interval $(]-L/2, +L/2[)^3$, and the boundaries consist of the closure of this domain. The velocity vector has three components ${\bm u} = (u_x, u_y, u_z)$, and the space positions are given by ${\bm x} = {x, y, z}$. A Dirichlet boundary condition for the velocity is enforced on the boundaries, consisting of a no-slip condition on all walls except on the moving lid at $x = +L/2$. The velocity on the lid is constant with a value $\bm{u}(x=+L/2, y, z) = (0, U, 0)$. It should be pointed that this problem is mathematically not strictly well defined given the velocity discontinuities in the edges and corners of the lid. For this reason, a smoothly increasing profile is sometimes defined in these regions. 
This issue is however ignored in the present study because, as the subsequent results show, the LBM copes with this discontinuity without problem.
The Reynolds number is defined as $\mathrm{Re} = U L /\nu$.
For illustrative purposes, the flow field obtained by our solver in a steady and in a turbulent regime is shown in Figure~\ref{fig:cavity-velocity-field}.

The problem is numerically solved with a homogeneous mesh of $N \times N \times N$ nodes, which includes a wall node on all extremities of the domain. Thus, the cell spacing is $\delta x = L / (N-2)$. The time discretization was fixed by setting the reference lattice velocity $u_\mathrm{LB} \equiv U (\delta t / \delta x)$ to a constant value of $0.06$. The Mach number, which in the present context of incompressible flow has a purely numerical significance, is therefore of $\mathrm{Ma}= 0.06 \sqrt{3}\approx 0.1$. The discrete time step is consequently defined as $\delta t =  (u_\mathrm{LB}/ U) \delta x$. The dimensionless time elapsed after $N_\mathrm{iter}$ iterations is $t = N_\mathrm{iter} \delta t (U/L) = N_\mathrm{iter} t_c$, with the dimensionless characteristic time being $t_c = \delta t (U/L)=u_\mathrm{LB} / (N-2)$.

In spite of the simple setup, this problem poses some challenges due to the velocity discontinuity between the top lid and lateral walls, and because of the complexity of the flow field and boundary layers that develop at higher Reynolds numbers.
This case is sometimes argued to pose substantial issues of accuracy and numerical stability for the LB method, which are for example overcome in~\cite{hegele_high-reynolds-number_2018} with help of a sophisticated boundary condition scheme. 
The results in this section show however that the LB framework of the \texttt{stlbm} project is sufficient to achieve satisfying results in all regimes, using either the standard BGK-W2 or the RR collision model.
%in the former case, and the Recursive-Regularized model (without additional Smagorinsky-like subgrid-scale modeling~\cite{SAGAUT_CMA_59_2010}) in the latter case.}

\begin{figure}[!hbt]
%\begin{center}
%\includegraphics[width=.45\textwidth]{re1000.pdf}
%\quad
%\includegraphics[width=.45\textwidth]{re3200.pdf}
%\end{center}
%\caption{{\bf Profile of cubic cavity flow at Re=1000 (left) and Re=3200 (right).}
\centering
\includegraphics[width=.99\textwidth]{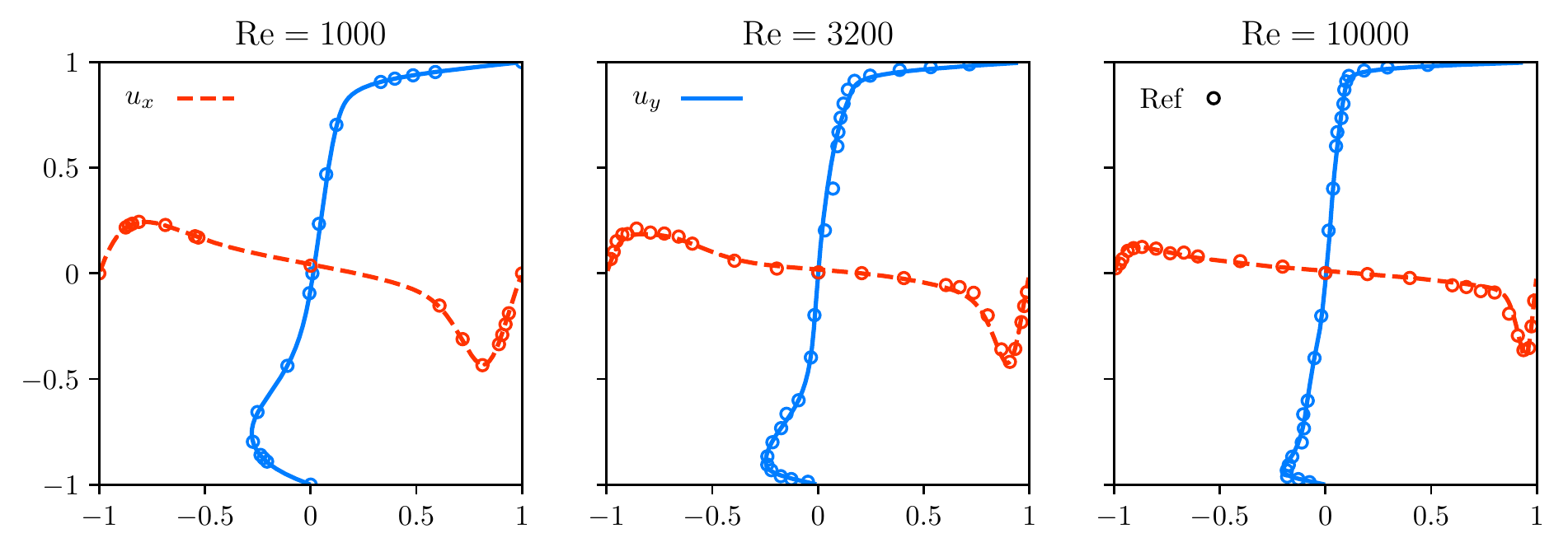}
\caption{{\bf Profile of cubic cavity flow for two Reynolds numbers.}
    Simulation results (solid lines) in a cubic lid-driven cavity along two center lines. Left image: $\mathrm{Re}=1000$, resolution $N=256$, collision model BGK-W2, comparison of the results with reference values (circles) from spectral simulations provided in~\cite{albensoeder_accurate_2005}. Middle image: non-stationary laminar flow at $\mathrm{Re}=3200$ simulated with a resolution $N=256$ and collision model BGK-W2, averaged between a dimensionless time $t=50$ and $t=250$. Right image: turbulent flow at $\mathrm{Re=10000}$ simulated with a resolution $N=400$ and RR collision model, averaged between a dimensionless time $t=150$ and $t=500$. The curves of the middle and right image are compared with experimental data extracted from figures in~\cite{prasad_reynolds_1989}.}
\label{fig:laminar-cavity}
\end{figure}

Figure~\ref{fig:laminar-cavity} plots the $y$-component of the velocity $u_y$ along the center line $y=z=0$ against the non-dimensionalized coordinate $2x/L$, and the $x$-component of the velocity $u_x$ along the center line $x=z=0$ against the coordinate $2y/L$.  On the left image, at $\mathrm{Re}=1000$, the flow is steady. The flow is measured at an instantaneous time step after reaching stationary state and compared against simulation data of a spectral code. On the middle image, at $\mathrm{Re}=3200$ the flow is laminar and unsteady. The profile is averaged between dimensionless time values of $t=50$ and $t=250$ and is compared against experimental data published in~\cite{prasad_reynolds_1989}. On the right image, at $\mathrm{Re}=10000$ the flow is turbulent. The profile is averaged between dimensionless time values of $t=150$ and $t=500$ and is again compared with~\cite{prasad_reynolds_1989}. Excellent agreements are achieved in all three cases.

% Place figure captions after the first paragraph in which they are cited.

%\bcol{Figure~\ref{fig:turbulent-cavity} plots the same data for a turbulent flow in a cavity at $\mathrm{Re}=10'000$, again compared with the experimental reference data of~\cite{prasad_reynolds_1989}. In this case, the profiles are averaged over a non-dimensionless time $t=100$ and $t=500$.}

\subsection{Parallel performance on different platforms}\label{sec:parallel-performance}
The case of a lid-driven cavity has been executed on different platforms to compare the performances. The \lstinline{for_each} loops over the cells were executed with the \lstinline{par_unseq} execution policy. 
%Furthermore, the indicated values are a median over 30 measurements, distributed over three days. 
Furthermore, reported values correspond to a median computed over 30 measurements, the latter being distributed other three days to reduce the impact on the benchmark of potential special events in the tested hardware or operating system. 
Each measurement consisted of 1000 warm-up iterations without time measurement, followed by 1000 benchmarked time steps.
Finally, all test cases were first executed with a BGK-W2 collision model at a resolution of $N=128$. A comparative study of the impact of the collision model on performances is also proposed in Section~\ref{subsubsec:coll_model_perfo}. 
As conventional in the field of LB, the performance is measured in terms of million lattice node updates per second (MLups), a value that identifies the number of lattice nodes, divided by one million, that are driven through a collision-streaming cycle in a second. 
To transpose this measurement in terms of time required to update one grid point per iteration (which is more commonly used in the Computational Fluid Dynamics community~\cite{MANOHA_AIAA_2846_2015}), one simply needs to take the inverse of the number of lattice node update per second. As an example, a performance of 10 MLups corresponds to $0.1\, \mu s/pt/ite$.

\begin{figure}[!hbt]
\begin{center}
\includegraphics[width=.6\textwidth]{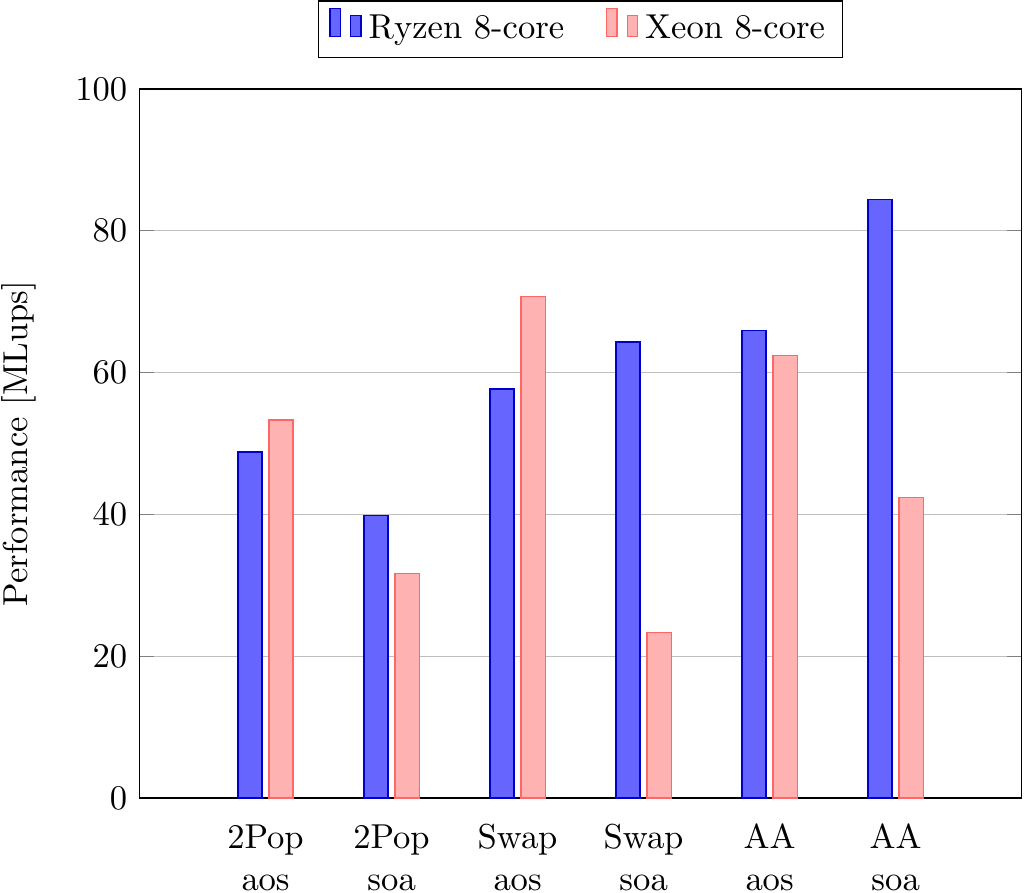}
\end{center}
\caption{{\bf Performance of CPUs with a modest core count.}
All simulations are executed with the BGK-W2 collision model and $N=128$. On both CPUs, best performance was quite unpredictably obtained sometimes with and sometimes without loop unrolling. The plotted value shows therefore always the best of the two.}
\label{fig:medium-core-count}
\end{figure}

The benchmark test was first executed on conventional CPUs with a modest core count. Figure~\ref{fig:medium-core-count} %low-core-count} 
shows the performance obtained using the six different data structures on a 8-core AMD and a 8-core Intel CPU. 
The results should not be used to set up a competitive comparison between the involved CPU models, which differ in the year of market release, the clock rate, and other parameters. All in all, the results show overall comparable performance across models. The Intel CPU yields consistently best performances with an %array-of-structure
aos layout, while best performances on the AMD CPU are most frequently obtained with the %structure-of-array 
soa approach, in agreement with the many-core AMD and the GPU architectures presented further down. This difference is mostly explained by the fact that the multi-core scaling performance of the code depends on the CPU and on the data structure. As shown below, the aos implementation strategy has better single-core performance than soa across architectures, but loses its edge over soa at a larger core count. While the AMD CPU reaches its best results with the AA-pattern, it is interesting to note that the swap-aos approach is the optimal choice on the Intel CPU, in spite of its apparent disadvantage due to the separated collision and streaming steps.

\begin{figure}[!bt]
\begin{center}
\includegraphics[width=.8\textwidth]{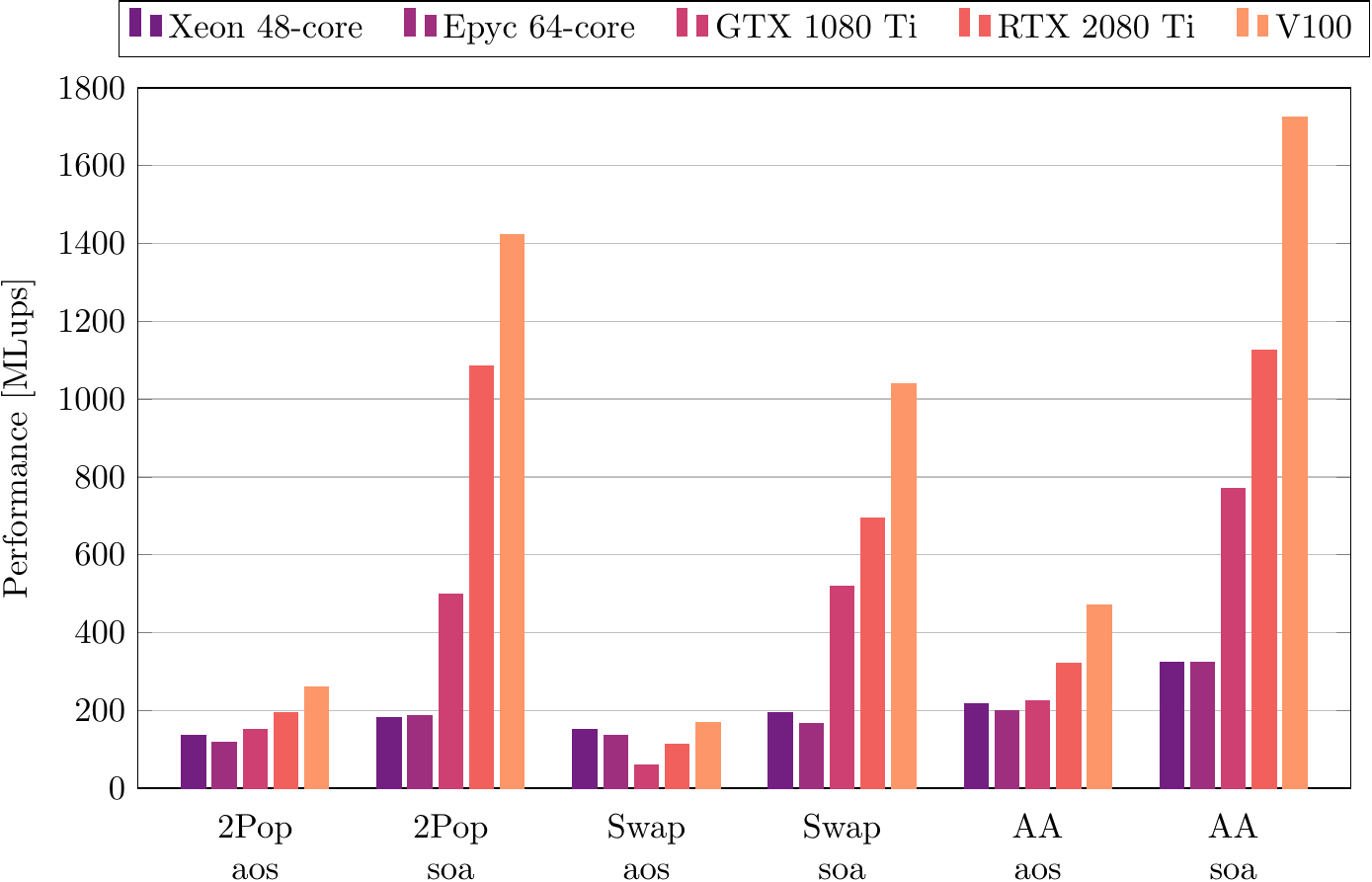}
\end{center}
\caption{{\bf Performance of many-core CPU and GPU architectures.}
All simulations are executed with BGK-W2 collision model and $N=128$. For best overall performance, the simulations on the 64-core Epyc CPU were executed without loop unrolling, while the 48-core Intel CPU and the two GPUs used loop unrolling.}
\label{fig:performance-manycore}
\end{figure}

In Figure~\ref{fig:performance-manycore}, the performances obtained on five many-core platforms are compared, namely a 64-core AMD EPYC, a 48-core Intel Xeon, two consumer-class GPUs and a data-center GPU. 
It is not typical to include GPUs and CPUs on a same comparison plot, as CPUs are often shown to perform one or two orders of magnitude worse than GPUs. 
In this case, the common comparison is motivated by the fact that they are measured with the exact identical code, but also by the fact that in our comparison, and if the best performing implementation scheme is selected for each platform, the faster of the two consumer-class GPUs outpaces the CPUs by only roughly a factor four. 
To achieve best performance, the loops were unrolled on the GPUs and on the Intel Xeon, but were used without unrolling on the AMD CPU. Best performances are systematically obtained with the soa data layout, and the AA-pattern is the clear winner across all CPU and GPU platforms.
For this reason, and also because of its lower memory needs, the AA-pattern (combined with the soa data layout) can be considered the most generally well performing choice for many-core platforms.

The real life implications of the performance metrics of the code are illustrated for the case of the turbulent cavity flow, which is run with the RR collision model at $N=400$ up to a dimensionless time of $t=500$ (to gather sufficient statistics), thus requiring a total of 3.3 million iterations. The simulation does not make use of any subgrid-scale turbulence model, but it is stabilized by adjusting the relaxation parameter associated to the bulk viscosity to a value of 1. On the 48-core Intel Xeon CPU, the execution time without outputs is of 7 days and 15 hours, while the RTX 2080 Ti completes the same task in 2 day and 15 hours, and the V100 PCIE in just 1 day and 3 hours.

Table~\ref{tab:speedup-cpu} shows the numerical values of the performances obtained on the two CPUs with the aos and the soa implementation. The performances obtained if the program uses only a single core are also shown. It can be seen that on both CPUs, the aos implementation is more performant on a single core, but the soa approach is preferable when all cores are exploited. It can be hypothesized that although in general, the aos approach makes better use of memory caching mechanisms, this advantage is lost when multiple cores compete for access to shared memory space, and cache coherence must be guaranteed. It is further remarked that both CPUs exhibit a comparable parallel efficiency, and that the Intel CPU compensates the lower core count with a stronger single-core performance.
\begin{table}[!htb]
    \centering
    \caption{{\bf Performance and speedup of AMD and Intel CPU with large core counts.} Detailed performance figures (in MLups) are provided for the cavity benchmark case with $N=128$, BGK collision, and AA-pattern implementation, for the 64-core AMD EPYC CPU and the 48-core Intel Xeon CPU, for program executions using 1 core or all cores. The speedup is the ratio between these two values, and the efficiency is a measure of the scaling performance (100\% efficiency stands for an ideal speedup).}
\begin{tabular}{|r+ccc|}
%\begin{tabular}{|r+SSS[table-format=4.1]|}

\hline
                          &  {\bf Performance} & {\bf Performance} & {\bf Speedup}\\
                          &  {\bf of 1 core} & {\bf of full CPU} & {\bf (Efficiency)}\\\thickhline
    64-core Epyc / AA aos &  11.68          & 199.0            & 17.0 (26.6\%)\\\hline
    64-core Epyc / AA soa &  7.42           & 322.7            & 43.5 (68.0\%)\\\hline
    48-core Xeon / AA aos &  12.32          & 216.9            & 17.6 (31.9\%)\\\hline
    48-core Xeon / AA soa &  10.13          & 323.7            & 36.7 (66.6\%)\\\hline
\end{tabular}
\label{tab:speedup-cpu}
\end{table}

Table~\ref{tab:numeric-values} provides the numerical values of the performances obtained on the two GPUs with the double-population implementation and the AA-pattern (soa implementation) and compares them with a reference performance obtained on the same GPUs with a code extracted from the CuBoltz code, an open-source library that evolved from the work described in~\cite{beny_efficient_2018}.
Written in the Cuda language, this code is written by one of the authors of the article and is carefully optimized for performance.
The extracted code has been reshaped to solve the exact same problem as the present \texttt{stlbm} tests, as verified through comparative regression tests.
The comparison shows that the generic \texttt{stlbm} code ends up only $21\%$ (GTX 1080 Ti) respectively $17\%$ (RTX 2080 Ti) below the performance of a hardware-specific, carefully optimized code. 
It can be concluded that the formalism of C++ Parallel Algorithms allows to achieve performances adequate for high performance computing, while at the same time offering an elegant and generic programming style.
\begin{table}[!tb]
\centering
\caption{{\bf Performance values of \texttt{stlbm} and the reference Cuda code CuBoltz} for the lid-driven cavity at $N=128$, with the BGK-W2 collision model. Performance is indicated in MLups, followed in parenthesis by the ratio between this performance and the theoretical peak performance, computed for a memory bandwidth of 480 GB/s (GTX 1080 Ti), 616 GB/s (RTX 2080 Ti), and 900 GB/s (V100 PCIE).}
\begin{tabular}{|r+ccc|}
%\begin{tabular}{|r+SSS[table-format=4.1]|}
\hline
                & {\bf CuBoltz}     & {\bf\texttt{stlbm} (2Pop)} & {\bf\texttt{stlbm} (AA)}\\\thickhline
    GTX 1080 Ti & 877.7 (56\%)  & 499.7 (32\%)  & 770.1 (49\%) \\\hline
    RTX 2080 Ti & 1415.8 (70\%) & 1084.9 (54\%) & 1124.8 (56\%) \\\hline
    V100 PCIE & 1823.6 (62\%) & 1422.6 (48\%) & 1723.7 (58\%) \\\hline
\end{tabular}
\label{tab:numeric-values}
\end{table}

On GPUs, it is useful to compare these performance values against a theoretical peak performance obtained under the assumption that computations are negligible and the overall performance is limited solely by optimally coalesced accesses to central memory. With the two-population scheme and the AA-pattern, all populations need to be read and written once per iteration, which in double-precision representation requires two memory transactions of an 8-byte floating-point number per population and iteration (with the Swap algorithm, the number of memory transactions is doubled). For these two implementation schemes, the peak performance $p$ (measured in LUPS) is therefore established by the formula
\begin{equation}
    p = bw / (2 \cdot 19 \cdot 8),
\end{equation}
where $bw$ is the memory bandwidth (in bytes per second). On table~\ref{tab:numeric-values}, the ratio of measured performance over the theoretical peak performance is indicated in parentheses. It can be seen that the \texttt{stlbm} code reaches performances close to, and in some cases even distinctly superior to, $50\%$ of the theoretical peak performance that can be expected on the respective GPUs. It is worth noting that these performances are obtained for lightly loaded GPUs, i.e., $N=128$. By further increasing the load to $N=400$, e.g., for the data-center GPU V100 PCIE, the ratio of measured performance is increased to $84\%$, $63\%$ and $72\%$ for CuBoltz, the two-population scheme and the AA pattern respectively. In any case, this study demonstrates both the quality of the stlbm algorithm, and the suitability of LB codes for GPU platforms.

\subsection{Analysis of performance-impacting features}

\subsubsection{Loop unrolling}
\begin{figure}[!bt]
\begin{center}
\includegraphics[width=.7\textwidth]{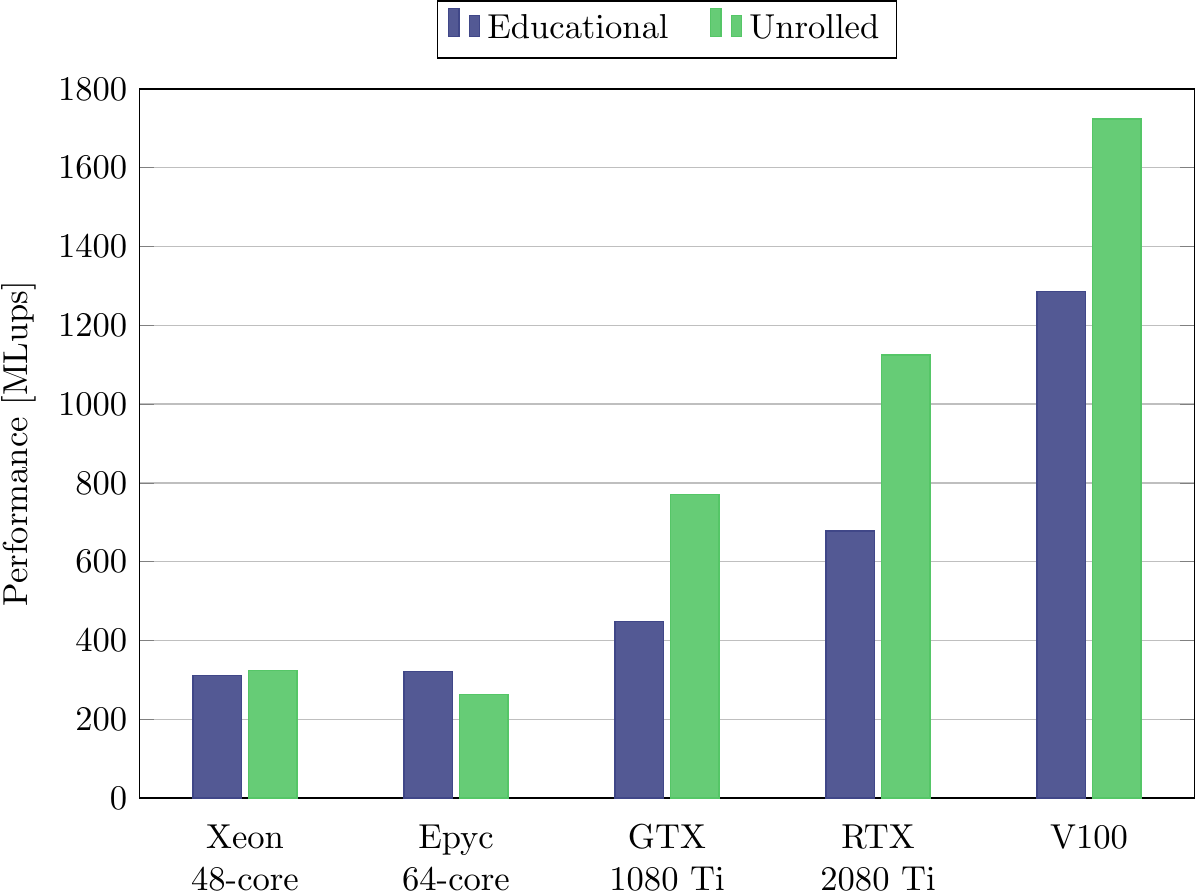}
\end{center}
\caption{{\bf Influence of loop unrolling.}
All simulations are executed with the BGK-W2 collision model, with the AA-soa scheme, and with $N=128$.}
\label{fig:unrolling}
\end{figure}
Figure~\ref{fig:unrolling} compares the performance of the educational and the unrolled codes on the five tested many-core platforms for the BGK-W2 collision model.
The results show that the GPUs experience substantial speed improvements from loop unrolling (the improvement is of $72\%$, $66\%$, and $34\%$ on the GTX 1080 Ti, the RTX 2080 Ti, and the V100 PCIE respectively), while the AMD EPYC performs better with the educational code, which presumably allows more targeted compiler optimizations. 
Interestingly, when the educational code is used, and thus, readability is emphasized over hand-tuned optimizations, the consumer-class GPUs seem to lose their clear-cut performance edge over CPUs, as they are only $39\%$ (GTX 1080 Ti) respectively $110\%$ (RTX 2080 Ti) faster than the AMD CPU. %, leading to the impression that GPUs may loose their clear-cut performance edge over CPUs. 
The data-center GPU V100 PCIE on the other hand remains substantially faster even for this version of the code, with a speed up of $298\%$ over the AMD.

\subsubsection{Collision model\label{subsubsec:coll_model_perfo}}
Hereafter, the performance obtained with different LB collision models are investigated on the fastest consumer-level platform, the RTX 2080 Ti GPU. In all cases, the optimized codes with loop unrolling were used, and multi-relaxation-time formulations of collision models were considered.   

\begin{figure}[!h]
\begin{center}
\includegraphics[width=.7\textwidth]{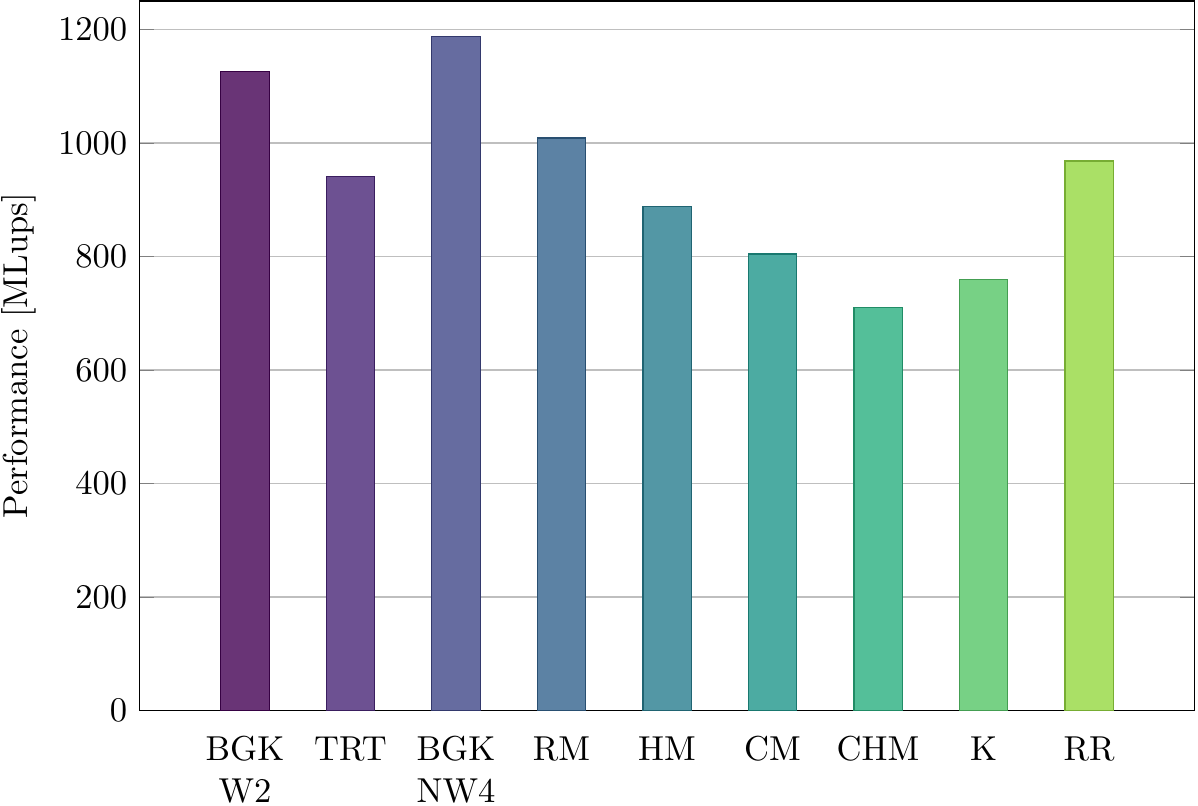}
\end{center}
\caption{{\bf Performance of different LB models.}
All simulations are executed with the AA-soa data structure at $N=128$ on the RTX 2080 Ti GPU.}
\label{fig:model-dependence}
\end{figure}

Generally speaking, Figure~\ref{fig:model-dependence} shows that the speed difference between the slowest and the fastest collision model remains within a factor two. 
This means that adopting modern collision models, e.g. to simulate high-Reynolds number flows in a stable and accurate manner, does not come at the expense of a drastic performance loss. 
Hence, LBMs based on these collision models remain particularly competitive as compared to Navier-Stokes solvers.
Quantitatively speaking, single-relaxation-time formulations (BGK-W2 and BGK-NW4) are the fastest. 
For the latter model, more redundant terms can be pre-computed, hence its better performance. 
Regarding collision models based on multiple relaxation times, it seems clear that increasing the number of moment space transitions automatically increases the number of floating point operations, hence a decrease of performance. 
As an example, the CHM-LBM requires two intermediate moment transformations before computing post-collision populations from their RMs, i.e. transformations from CHMs to HMs and from HMs to RMs. 
The same goes for the K-LBM, whereas the HM-LBM only requires a single moment transformation (from HMs to RMs). %, even though it might be less stable depending on the values of high-order relaxation times~\cite{RSTA}. 
In addition, even if the RR-LBM is also based on a collision applied in the HM space, it is faster than the HM-LBM. 
This is explained by the fact that it only requires the computation of second-order HMs, and recursive computations of high-order HMs only add a few floating point operations.

In the end, the present study confirms that the good performances obtained with the \texttt{stlbm} implementation strategy are consistent across all collision models.

\subsubsection{Domain size}
\begin{figure}[!h]
%\begin{center}
%\includegraphics[width=.45\textwidth]{24sept_BGK.pdf}
%\quad
%\includegraphics[width=.45\textwidth]{24sept_RR.pdf}
%\end{center}
%\caption{{\bf Dependence of domain size on performance on GPU.} Left: BGK-W2; Right: RR. Simulations are run on a RTX 2080 Ti GPU.}
\centering
\includegraphics[width=.9\textwidth]{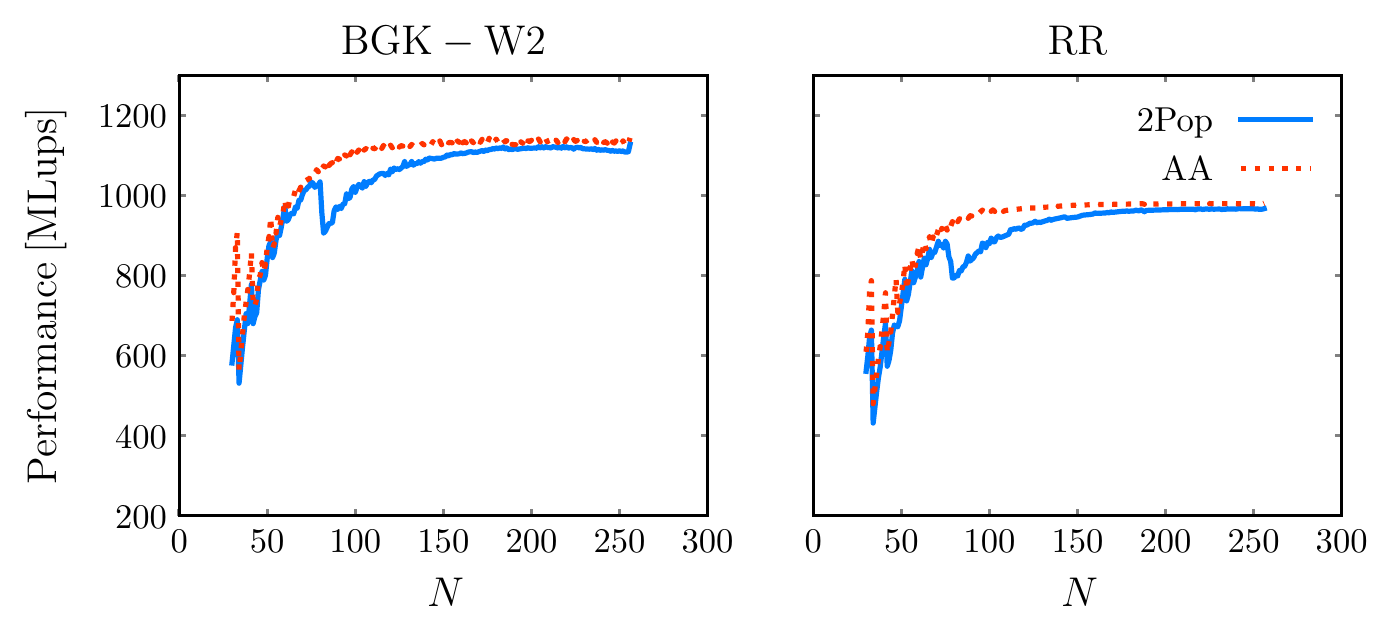}
\caption{{\bf Dependence of domain size on performance on GPU.} Left: BGK-W2; Right: RR. Simulations are run on an RTX 2080 Ti GPU.}
\label{fig:domain-size}
\end{figure}

Figure~\ref{fig:domain-size} shows the performance achieved by \texttt{stlbm} on the fastest consumer-level hardware platform, the RTX 2080 Ti GPU, using different domain sizes, with increments of 1 for the number of cells $N$ in every direction. 
The performance figures are shown for the AA-soa and double-population-soa implementation strategies, and for the BGK-W2 and RR collision models,
to test both a simple and a complex model.
%\bcol{that are used in the validation section, Sec~\ref{sec:validation} (possible update depending on results at Re=10'000)}. 
In general, the performance increases up to a resolution of roughly $100\times 100\times 100$, and stays roughly constant afterwards.
The double-population implementation further shows a drop of performance close to $N=90$, which is not further explained at this point. 
This investigation finally leads to the conclusion that the performances of the double-population scheme and the AA-pattern are very similar, and that the slight superiority of the double-population approach observed in Sec~\ref{sec:parallel-performance} would be less clear-cut or even inexistent at a different domain size or with a different collision model.

%\subsubsection{Domain ratio}

%\subsubsection{Vectorization}

\section{Conclusion\label{sec:conclusion}}
This article presents an implementation approach for multi-threaded LB applications which is hardware agnostic, relies exclusively on standard C++ language features, yet exhibits very interesting performances on many-core CPUs and on GPUs. 
It is presented and disseminated in form of the open-source C++ library \texttt{stlbm}. 
While the computational kernels are kept very simple to serve as templates in general-purpose LB codes, and use a descriptive programming style to serve an educational purpose, the \texttt{stlbm} library targets a broad scope, as it offers six different implementation strategies, and nine LB collision models, which can be combined without restrictions. 

The article presents performance metrics of the code for different implementation choices and on different platforms for the test case of a 3D lid-driven cavity flow. The results of these benchmarks show that the proposed framework leads to excellent performance on both CPUs and GPUs. In the latter case, very similar performances are obtained compared to those of an optimized code written in a domain-specific language, with a performance loss of only approximately 20\%. 

The results of these benchmarks also lead to an interesting conclusion that appears to question common prejudice regarding the use of CPUs vs. GPUs in computational science. 
While it is traditionally thought that GPU development is substantially more time consuming than CPU development, but more rewarding thanks to performance gains of one to two orders of magnitude, the \texttt{stlbm} library presented in this article paints rather the opposite picture. Indeed, the same effort leads to a code that compiles to a CPU and a GPU binary without changes. As for the performance, the two consumer-level GPUs (which, like the CPUs, were deployed within desktop computers) outperform the CPUs by roughly a factor 4 (or a factor 2 if the more readable version of the code is considered), while the data-center GPU achieves a factor of more than 5 (or a factor 4 for the educational code version).

The achieved performances are barely affected by the use of sophisticated collision terms.
More precisely, the speed difference between standard and more sophisticated collision models remains within a factor two, which means the most robust collision models remain competitive (despite their increased complexity) as compared to Navier-Stokes solvers.

The numerical tests also allow to single out two implementation strategies that exhibit the best performances across the tested high-performance platforms, namely the double-population scheme and the AA-pattern, both used in combination with a structure-of-array data layout. Among these two, the AA-pattern appears to be more versatile thanks to its consistently good performances across different platforms. It furthermore stands out because of its reduced memory needs. While the third tested approach, the swap scheme, performed poorly on GPU and AMD CPU hardware, it produced good performance on the 8-core Intel CPU, combined with an array-of-structure layout. We therefore retain also this approach as an interesting candidate, which has a low memory footprint like the AA-pattern. It furthermore proposes an algorithm that fits more easily into complex LB applications, as it does not require a separation into even and odd time steps.

As a consequence of these observations, the \texttt{stlbm} library offers, additionally to its full software framework, three standalone, single-file codes based on the two-population soa, the AA-pattern soa, and the swap aos strategies respectively. Each of these standalone files contains less than 500 lines of code and allows to understand the philosophy of LB simulations with Parallel Algorithms with fewer hurdles than the full code basis. They are written for the BGK-W2 collision model, which is however easily replaced by any of the other models by inserting corresponding code portions for the \texttt{stlbm} library.

We finally point out that the conclusions regarding the best performing implementation scheme appear not to be universal, and rather linked to the specificity of the presently tested hardware. The generic version of the \texttt{stlbm} code serves therefore an important purpose, allowing all implementation strategies to be rapidly tested on future many-core systems.

\section{Supporting information}\label{sec:supporting_information}
\paragraph*{S1 Source Code.}
\label{S1_Source Code}
{\bf STLBM software library.} The \texttt{stlbm} software library is released as an open-source project and can be used under the terms of an MIT license. The source code is available at \url{https://gitlab.com/UniGeHPFS/stlbm}. All performance measurements presented in this article were executed with the version of the code identified through the Git tag \texttt{benchmarks\_plosone}.

\paragraph*{S1 File.}
\label{S1_File}
{\bf Performance Measurements.xlsx.}  This spreadsheet contains the detailed performance measurements used to produce the median values on the graphs and in the tables of this article. 

\section{Acknowledgments}
We would like to thank Orestis Malaspinas for useful discussions regarding the role of functional programming styles (in relationship with Futhark) in hardware-agnostic multi-threaded program development, and for supplying testing hardware for this article. We would further like to thank Christos Kotsalos and Anthony Boulmier for providing advice and help on hardware installation and configuration. We finally acknowledge funding of one of the three authors by
the Swiss PASC project ``An HPC framework for blood
flow simulations in vasculature and in medical devices''.

%\nolinenumbers

%\bibliography{stlbm}
% Either type in your references using
% \begin{thebibliography}{}
% \bibitem{}
% Text
% \end{thebibliography}
%
% or
%
% Compile your BiBTeX database using our plos2015.bst
% style file and paste the contents of your .bbl file
% here. See http://journals.plos.org/plosone/s/latex for 
% step-by-step instructions.
% 
\bibliographystyle{unsrt}

\end{document}